\documentclass[12pt]{article}
\usepackage[a4paper, top=16mm, text={200mm, 248mm}, includehead, includefoot, hmarginratio=1:1, heightrounded]{geometry}
\usepackage{amsmath,amssymb,mathrsfs,amsthm,tikz,shuffle,paralist,float}
\usepackage{dsfont}
\usepackage{color}

\definecolor{darkred}{RGB}{173,34,48}
\usetikzlibrary{shapes.geometric}

\usetikzlibrary{snakes}
\usetikzlibrary{calc}
\usetikzlibrary{decorations}

\usepackage[all]{xypic}
\usepackage{jheppub}
\usepackage{hyperref}
\usepackage{subfig}
\usepackage{caption}
\usepackage{diagbox}
\usepackage{bm}
\usepackage{slashed}
\usepackage[normalem]{ulem}

\theoremstyle{plain}

\title{New relations for tree-level form factors and scattering amplitudes}

\author[a,c]{Jin Dong\footnote{dongjin@itp.ac.cn}}
\author[a,b, d]{Song He\footnote{songhe@itp.ac.cn}}
\author[a,e]{Guanda Lin\footnote{linguandak@pku.edu.cn}}
\affiliation[a]{
CAS Key Laboratory of Theoretical Physics, Institute of Theoretical Physics, Chinese Academy of Sciences, Beijing 100190, China}
\affiliation[b]{School of Fundamental Physics and Mathematical Sciences, Hangzhou Institute for Advanced Study, ICTP-AP, UCAS, Hangzhou 310024, China;\\
International Centre for Theoretical Physics Asia-Pacific, Beijing/Hangzhou, China}
\affiliation[c]{School of Physical Sciences, UCAS, No.19A Yuquan Road, Beijing 100049, China}
\affiliation[d]{Peng Huanwu Center for Fundamental Theory, Hefei, Anhui 230026, P. R. China}
\affiliation[e]{Higgs Centre for Theoretical Physics, University of Edinburgh,
James Clerk Maxwell Building, Peter Guthrie Tait Road, Edinburgh, EH9 3FD,
United Kingdom}
\date{\today}

\abstract{ 
We show that tree-level form factors with length-two operators in Yang-Mills-scalar (YMS) theory exhibit structures very similar to scattering amplitudes of gluons and scalars, which leads to new relations between them. Just like amplitudes, $n$-point Yang-Mills form factors with ${\rm tr}(F^2)$ operator can be decomposed as a linear combination of form factors with ${\rm tr}(\phi^2)$ operator and $r$ external scalars in YMS theory, where the coefficients are given by Lorentz products of the $r$ linearized field strengths. Moreover, we show that any such $n$-point form factor of ${\rm tr}(\phi^2)$ operator can be further expanded into $(n{+}1)$-point YMS amplitudes with an additional off-shell scalar leg. In addition to unravelling hidden structures, our results provide an efficient algorithm for computing all-multiplicity length-two form factors in any dimension, as well as their Cachazo-He-Yuan formulae via those of the YMS amplitudes. }

\begin{document}

\maketitle


\section{Introduction}
Recent years have witnessed enormous progress in unravelling hidden structures and especially relations for scattering amplitudes in QFT (c.f.~\cite{Elvang:2013cua}). 
For example, scattering amplitudes of gluons, gravitons, and Goldstone particles {\it etc.} are closely related to each other via a web of deep relations; perhaps the most famous ones are the double-copy relations (see the review~\cite{Bern:2019prr} and references therein), originally discovered as field-theory limit~\cite{Bjerrum-Bohr:2009ulz} of tree-level Kawai-Lewellen-Tye(KLT) relations in string theory~\cite{Kawai:1985xq} and extended to loop level by Bern-Carrasco-Johansson(BCJ) via color-kinematics duality in QFT~\cite{Bern:2008qj,Bern:2010ue}. Such tree-level double-copy relations have been extended to a large class of theories including EFTs {\it e.g.} using Cachazo-He-Yuan (CHY) formulae~\cite{CHY1,CHY2,CHY3,CHY5,CHY4}, where they are naturally interpreted as a ``direct product" operation, for amplitudes in various theories. Another natural operation can be called ``direct sum", which produce mixed amplitudes of {\it e.g.} gravitons and gluons in Einstein-Yang-Mills theory (EYM) or those in Yang-Mills-scalar (YMs). More recently, a different kind of amplitude relations~\cite{Dong:2021qai}, dubbed ``universal expansions" have been shown using the uniqueness theorem in~\cite{Arkani-Hamed:2016rak,Rodina:2016jyz,Rodina:2016mbk} in all these theories. Originally discovered using CHY formulas~\cite{Lam:2016tlk,Fu:2017uzt,Du:2017kpo}, such relations expand {\it e.g.} gravity (gluon) amplitude into linear combination of those in EYM (YMs) amplitudes, and in a sense they interpolate between the two operations above. 

On the other hand, as the simplest off-shell extension of amplitudes, form factors have also attracted increasing attentions recently since numerous  structures have been uncovered especially at multi-loop level in ${\cal N}=4$ SYM theory ({\it c.f.} ~\cite{Brandhuber:2012vm,Brandhuber:2014ica,Loebbert:2015ova,Loebbert:2016xkw,Brandhuber:2017bkg,Brandhuber:2018xzk,Sever:2020jjx,Dixon:2020bbt,Dixon:2021tdw,Dixon:2022rse,Sever:2021nsq,Sever:2021xga,Boels:2012ew,Yang:2016ear,Lin:2020dyj, Lin:2021kht,Lin:2021qol,Lin:2021lqo}).
Despite all these developments, it is fair to say that much less is known even for tree-level form factors in gauge theories in general dimensions. For example, while all-multiplicity BCJ expressions are known for tree amplitudes in any dimension~\cite{Fu:2017uzt,Teng:2017tbo,Du:2017kpo, Edison:2020ehu, He:2021lro, Cheung:2021zvb}, no such results or even evidence for CK duality has been found for $n$-point Yang-Mills form factors with {\it e.g.}  ${\rm tr}(F^2)$ operator. To our best knowledge, such form factors in general dimension are still computed mainly using Feynman diagrams, which quickly get out of control. Relatedly, mainly due to the off-shellness of the operator in form factors, it is much more difficult to apply string-inspired methods such as CHY formulas to form factors (see~\cite{He:2016jdg,Brandhuber:2016xue,He:2016dol} for such a twistor-string formula in four dimensions). It is thus natural to ask if any of the above structures for gluon amplitudes can be found in form factors, and if one could connect form factors to amplitudes in some way.

In this paper, we make a first step in answering these questions by proposing a natural decomposition of length-two form factors into scattering amplitudes in Yang-Mills-scalar (YMS) theory~\cite{Chiodaroli:2014xia, CHY4}, which in turn provide an efficient way for computing form factors explicitly. Our main results can be divided into two parts. First, in sec.~\ref{sec:F2decomp} we present a nice decomposition of Yang-Mills form factors with operator ${\rm tr}(F^2)$ (which carries  off-shell momentum $q$)
\begin{equation}
    F_n^{\operatorname{tr}(F^2)}(1,\ldots,n)=\int {\rm d}^{D}x \ e^{iqx} \langle 0|\operatorname{tr}(F^2)|1^{g},\ldots,n^{g}\rangle\,,
\end{equation}
and it is very similar to the expansion of Yang-Mills amplitude into a linear combination of YMS ones. More precisely, we show that the $F_n^{{\rm tr}(F^2)}$ can be written as a linear combination of $F_n^{{\rm tr}(\phi^2)}$ with $r$ scalars and $n{-}r$ gluons for $r=2, 3, \cdots, n$, and the coefficients are given by Lorentz trace of $m$ linearized field strengths. 

Moreover, in sec.~\ref{sec:phi2decomp} we propose that these YMS form factors $F_n^{{\rm tr}(\phi^2)}$ in turn can be decomposed into a sum of $(n{+}1)$-point YMS amplitudes with one additional (off-shell) scalar leg. Note that the latter has $r{+}1$ bi-adjoint scalars 
, and in principle such an expansion is not too surprising since their Feynman diagrams take similar form. To illustrate this point, let us consider the simplest ${\rm tr}(\phi^2)$ form factor, $F^{\rm tr(\phi^2)}_n (1^\phi, 2^g, \cdots, (n{-}1)^g, n^\phi)$ where the $2$ scalars are chosen to be adjacent in the ordering. All contributing Feynman diagrams have a scalar line connecting $1$ and $n$, and it is obvious that the operator, or ``$q$-leg", must be inserted via a $\phi^3$ vertex along the line. These Feynman diagrams are identical to all those for the color-ordered amplitude with $n{-}2$ gluons coupled to (adjacent) scalars, $1, n, n{+}1{=}q$, $A^{\rm YMS} (1^\phi, 2^g, \cdots, (n{-}1)^g, n^\phi, (n{+}1)^\phi)$. More precisely, since $q^2\neq 0$ for form factors but the YMS amplitudes are defined for massless momenta (in any dimension), the equality holds on the support of momentum conservation with the following prescription for the off-shell leg $q\to -\sum_{i=1}^n p_i$:
\begin{equation} \label{eq:prototype}
F_n^{{\rm tr}(\phi^2)} (1^\phi,\cdots, n^\phi)=A_{n{+}1}^{\rm YMS} (1^\phi, \cdots, n^\phi, q^{\phi})|_{q\to -\sum_{i=1}^n p_i}\,,
\end{equation}
where we have suppressed the gluons, and it is crucial that we need to express both sides in terms of on-shell momenta $p_1, \cdots, p_n$ only. Among other things, the identification with YMS amplitudes provides one explanation why such special form factors can be arranged to satisfy color-kinematics duality and double copy to interesting quantities for gravity~\cite{Lin:2021pne}. We will see that \eqref{eq:prototype} is a prototype for the new relations which expresses general $n$-point form factor into $n{+}1$-point amplitudes, and throughout the paper such identities are always understood with this prescription. 

In sec.~\ref{sec:expand and CHY}, by combining these two types of expansions, we obtain $F_n^{{\rm tr}(F^2)}$ as a linear combination of $n{+}1$-point YMS amplitudes, with coefficients given in terms of traces of field strengths. As a consequence, we have CHY formulae for all these form factors via CHY for YMS amplitudes, which reveal even more unexpected simplicity and give the first example of worldsheet formula for form factors in general dimension. Such formulae also provide a new method for explicitly computing all-multiplicity form factors (both for ${\rm tr}(\phi^2)$ and ${\rm tr}(F^2)$ operator), which are much more efficient than Feynman diagrams. 


\section{Decomposition of the ${\rm tr}(F^2)$ form factor into ${\rm tr}(\phi^2)$ ones} \label{sec:F2decomp}
The Yang-Mills-scalar amplitudes and Yang-Mills amplitudes have close relations. Physically, it is not hard to understand that by ``pulling out" polarizations, Yang-Mills amplitudes can become Yang-Mills-scalar amplitudes. In fact, this is also the case for form factors. In this section, after first reviewing the amplitudes relations, we are going to establish a parallel relation between pure Yang-Mills form factors and Yang-Mills-scalar form factors. 

Before writing down the exact relation, we review the definition of the theory and amplitudes/form factors that we are concerned about. 
We are considering the Yang-Mills-scalar theory with the Lagrangian \cite{Chiodaroli:2014xia}~\footnote{This theory can be obtained by coupling bi-adjoint $\phi^3$ with the dimensional reduction of Yang-Mills theory; in the literature the latter is sometimes denoted as YMS theory and the former generalized YMS theory, but we will only use YMS for the theory with coupling to bi-adjoint $\phi^3$.}: 
\begin{equation}\label{eq:ymsL}
\begin{aligned}
    \mathcal{L}^{\mathrm{YMS}{+}\phi^3}=&-\frac{1}{2} \operatorname{tr}_{\mathrm{C}}\left(D_{\mu} \Phi^{I} D^{\mu} \Phi^{I}\right)-\frac{1}{4} \operatorname{tr}_{\mathrm{C}}\left(F_{\mu \nu} F^{\mu \nu}\right)-\frac{g^{2}}{4} \operatorname{tr}_{\mathrm{C}}\left(\left[\Phi^{I}, \Phi^{J}\right]^{2}\right)\, \\
    &-\frac{\lambda}{3 !} f_{I, J, K} f_{\tilde{I}, \tilde{J}, \tilde{K}} \Phi^{I, \tilde{I}} \Phi^{J, \tilde{J}} \Phi^{K, \tilde{K}} \,, 
\end{aligned}
\end{equation}
where $\Phi^{I}=\sum_{a}\Phi^{I}_{\tilde{I}}T^{\tilde{I}}$ is the scalar field with flavor index $I$ and (a sum over) color index $\tilde{I}$, and $F^{\mu\nu}=F^{\mu\nu}_{\tilde{I}}T^{\tilde{I}}$ is the Yang-Mills field strength. Also, we denote the color and flavor trace by $\operatorname{tr}_{\rm C}$ and  $\operatorname{tr}_{\rm FL}$ respectively. 

Moreover, we are mostly interested in the single-trace double ordered amplitudes
\begin{equation}
    A^{\rm YMS}(i_1\ldots i_r|1\ldots n):=\langle 0| 1^{g}\ldots i_1^{\phi} \ldots,j^{g} \ldots i_r^{\phi} \ldots n^{g}\rangle \Big|_{\operatorname{tr}_{\rm C}(T^{1}\cdots T^{n}),\operatorname{tr}_{\rm FL}(Y^{i_1}\cdots Y^{i_r})}\,,
\end{equation}
and similarly the single-trace double ordered form factors
\begin{align}
    F^{\operatorname{tr}(\phi^2)}(i_1 \ldots i_r&|1 \ldots n)\\
    \nonumber &=\int d^{D}x\  e^{\mathrm{i}q\cdot x} \langle 0| \mathcal{O}_{\Phi} | 1^{g} \ldots i_1^{\phi} \ldots j^{g}\ldots i_r^{\phi} \ldots n^{g}\rangle \Big|_{\operatorname{tr}_{\rm C}(T^{1}\cdots T^{n}) \operatorname{tr}_{\rm FL}(Y^{i_1}\cdots Y^{i_r})}\,.
\end{align}
Here the operator is defined as $\mathcal{O}_{\Phi}=\operatorname{tr}(\phi^2):=\sum_{I,\tilde{I}} \Phi^{I,\tilde{I}}\Phi_{I,\tilde{I}}$. 

\subsection{Review of the decomposition for amplitudes}

The expansion of Yang-Mills amplitudes on Yang-Mills-scalar amplitudes was originally proposed in \cite{Lam:2016tlk,Fu:2017uzt,Du:2017kpo} by studying the CHY representation of Yang-Mills and Yang-Mills scalar amplitudes, and later categorized as a special case in a class of universal expansions \cite{Dong:2021qai}. The explicit formula is
\begin{equation}\label{eq:YMdecomp}
    \hskip -3pt
    A^{ \rm YM}(1,\ldots, n)=\sum_{r=0}^{n-2}\sum_{i_1<\ldots<i_r}\sum_{\sigma\in S_r} \mathrm{W}^{\rm f}_{1n}(\sigma(i_1, \ldots, i_r)) A^{ \rm YMS}(1,\sigma({i_1},\ldots,{i_r}), n|1,2\ldots, n)\,,
\end{equation}
where the``open trace" of the field strength ${\rm f}_{i}^{\mu\nu}=p_i^{\mu}\epsilon_{i}^{\nu}-p_i^{\nu}\epsilon_{i}^{\mu}$ is
\begin{equation}
    \mathrm{W}^{\rm f}_{1n}(j_1,\ldots,j_r):= \epsilon_{1,\mu_1} (\mathrm{f}_{j_1})^{\mu_1}_{\mu_2}\cdots (\mathrm{f}_{j_r})^{\mu_r}_{\mu_{r+1}} \epsilon_n^{\mu_{r+1}}\,.
\end{equation} 
To understand \eqref{eq:YMdecomp}, one should first notice that both particle $1$ and $n$ are special. For particles $2,\ldots,(n-1)$, the gauge invariance is manifest in \eqref{eq:YMdecomp}. The gauge invariance of particle $1$(or $n$), however, is not obvious. Actually, according to \cite{Arkani-Hamed:2016rak}, if we express the Yang-Mills amplitudes with a local form\footnote{Here the terminology "local form" only refers to cubic diagram expansion in this paper. In the original argument, the authors of \cite{Arkani-Hamed:2016rak} show that one can start from a more general ansatz.}, as is the case in \eqref{eq:YMdecomp}, it is only possible to manifest the gauge invariance of $(n-2)$ particles. 
Furthermore, as proven in \cite{Dong:2021qai}, requiring the gauge invariance of $(n-2)$ particles already spits out the $\mathrm{W}_{1n}^{\rm f}$ factor, yet one can not confirm that the  gauge invariant block accompanied with $\mathrm{W}_{1n}^{\rm f}$ is indeed a YMS amplitude. The next step is to use the gauge invariance of another particle say the first particle, which uniquely fixes the Yang-Mills amplitudes, as well as its expansion \eqref{eq:YMdecomp}. Note that to make this gauge invariance condition obvious, we have to combine all terms in the local expression such as \eqref{eq:YMdecomp} together  and reach a ``non-local" form with more poles in the denominator compared with cubic diagrams. For instance, for the four-gluon amplitude, we have (see \cite{Green:1987sp}, and we use the notations in \cite{Bern:2017tuc})
\begin{equation}
    A_4^{ \rm YM}=\frac{T_8}{st}, \text{ with } T_{8}=\frac{1}{2}\left(4\operatorname{tr}^{\rm f}(1,2,3,4)-\operatorname{tr}^{\rm f}(1,2)\operatorname{tr}^{\rm f}(3,4)\right)+ \text{cyclic}(2,3,4),
\end{equation}
which trivialized the gauge invariance of all polarizations. However, there are two Mandelstams in the denominator rather than one for four-point cubic diagrams. 

It is also helpful to understand the expansion \eqref{eq:YMdecomp} from the perspective of  transmuted operators~\cite{Cheung:2017ems}. By taking the following (ordered) ``open-trace" derivatives $\widetilde{\mathcal{D}}_{(i_1,i_2,\ldots, i_{r{-}1},i_r)}$
\begin{equation}
    \widetilde{\mathcal{D}}_{(i_1,i_2,\ldots, i_{r{-}1},i_r)}:= \partial_{({\color{red}\bm \epsilon_{i_{1}}}\cdot p_{i_2})}\partial_{(\epsilon_{i_2}\cdot p_{i_3})}\cdots \partial_{(\epsilon_{i_{r{-}1}}\cdot {\color{red}\bm \epsilon_{i_{r}}})}
\end{equation}
with the property  
\begin{equation}
\begin{aligned}
    &\widetilde{\mathcal{D}}_{(1,i_1, \ldots, i_r,n)} \mathrm{W}_{1n}(j_1, \ldots, j_r)=1 \Leftrightarrow (i_1, \ldots, i_r)=(j_1, \ldots, j_r)\,,\\
    &\widetilde{\mathcal{D}}_{(1,i_1, \ldots, i_r,n)} \mathrm{W}_{1n}(j_1, \ldots, j_m)=0 \text{ for any other situations}\,,
\end{aligned}
\end{equation}
one can extract the YMS amplitude from the YM one as 
\begin{equation}\label{eq:YMderivative}
    \widetilde{\mathcal{D}}_{(1,\sigma({i_1},\ldots,{i_r}),n)}A^{\rm YM}(1\ldots n)=A^{ \rm YMS}(1,\sigma({i_1},\ldots,{i_r}), n|1,2,\ldots, n)\,.
\end{equation}
This relation\footnote{Strictly, we need to choose a kinematic basis where $p_n$ is eliminated by momentum conservation for this relation to be true.} will also be useful when discussing the factorizations of form factor expansions in sec.~\ref{ssec:trF2proof}. 

Below we will see that  for form factors, a decomposition similar to \eqref{eq:YMdecomp} holds but manifests gauge invariance for all particles. Let us start from some simple examples. 

\subsection{Three- and Four-point examples for form factors}
Now we study the simple three- and four-point form factors to see the pattern for the form factor expansion.

\subsubsection{The three-point example}
Let us start from the simplest three-point form factor $F_{3}^{\operatorname{tr}(F^2)}$: one can reorganize the expression in the following form\footnote{Note that $\operatorname{tr}^{\rm f}(1,2)=2 (p_1\cdot \epsilon_2)(p_2\cdot \epsilon_1)-2(p_1\cdot p_2)( \epsilon_1\cdot \epsilon_2)$ has an overall factor 2 and $\operatorname{tr}^{\rm f}(1,2,3)=(p_1\cdot \epsilon_2)(p_2\cdot \epsilon_3)( p_3\cdot \epsilon_1)+\cdots$ does not.}
\begin{equation}\label{eq:3pttrF2}
\begin{aligned}
    F^{\operatorname{tr}(F^2)}_3(1^g,2^g,3^g)=&\operatorname{tr}^{\rm f}(1,2) \left(\frac{p_1\cdot \epsilon_3 }{s_{13}}+\frac{-p_2\cdot \epsilon_3}{s_{23}}\right)+\text{cyclic}(1,2,3)\\
    &+2 \operatorname{tr}^{\rm f}(1,2,3)\left(\frac{1}{s_{12}}+\frac{1}{s_{13}}+\frac{1}{s_{23}}\right)\, ,
\end{aligned}
\end{equation}
where $s_{i_1,i_2,\ldots,i_k}\equiv (p_{i_1}+p_{i_2}+\ldots +p_{i_k})^2$.The key feature in \eqref{eq:3pttrF2}, similar to the amplitudes, is to strip part of the polarizations off and leave gauge-invariant blocks containing the rest polarizations. These gauge-invariant blocks turn out to be scalar form factors:
\begin{equation} \label{eq:3pttrphi2}
\begin{aligned}
    &F^{\operatorname{tr}(\phi^2)}_3(1^\phi,2^\phi|1^\phi,2^\phi,3^g)=\left(\frac{p_1\cdot \epsilon_3 }{s_{13}}+\frac{-p_2\cdot \epsilon_3}{s_{23}}\right)\\
    &F^{\operatorname{tr}(\phi^2)}_3(1^\phi,2^\phi,3^\phi|1^\phi,2^\phi,3^\phi)=\left(\frac{1}{s_{12}}+\frac{1}{s_{13}}+\frac{1}{s_{23}}\right)
\end{aligned}
\end{equation}
so that
\begin{equation}\label{eq:3pttrF2b}
\begin{aligned}
    F^{\operatorname{tr}(F^2)}_3(1^g,2^g,3^g)=&\operatorname{tr}^{\rm f}(1,2) F^{\operatorname{tr}(\phi^2)}_3(1^\phi,2^\phi|1^\phi,2^\phi,3^g)+\text{cyclic}(1,2,3)\\
    &+ 2\operatorname{tr}^{\rm f}(1,2,3)F_3^{\operatorname{tr}(\phi^2)}(1^\phi,2^\phi,3^\phi|1^\phi,2^\phi,3^\phi)\,.
\end{aligned}
\end{equation}
From the three-point example, we see the basic difference between the form factor decomposition \eqref{eq:3pttrF2b} and the amplitudes ones are (1) instead of ``open-traces" in \eqref{eq:YMdecomp}, here we have ``closed-trace" defined as 
\begin{equation}
    \operatorname{tr}^{\rm f}(i_1,\ldots, i_r)=({\rm f}_{i_1})^{\mu_1}_{\mu_2}({\rm f}_{i_2})^{\mu_2}_{\mu_3}\cdots ({\rm f}_{i_r})^{\mu_r}_{\mu_1}\,;
\end{equation}
(2) \eqref{eq:3pttrF2b} is manifestly gauge invariant for all particles, in contrast to \eqref{eq:YMdecomp} for only $(n-2)$ of them; (3) Crossing symmetry is trivial for \eqref{eq:3pttrF2b}. 

\subsubsection{The four-point example}
Then we turn to the four-point example. Given the three properties mentioned in the three-point example, we are expecting an expansion looking like 
\begin{align}\label{eq:4pttrF22}
     F_4^{\operatorname{tr}(F^2)}(1,2,3,4)&
     =
     \lambda_2 \sum_{i_1<i_2}\operatorname{tr}^{\rm f}(i_1,i_2) F_4^{\operatorname{tr}(\phi^2)}(i_1,i_2|1,2,3,4) \nonumber\\
     &+\lambda_3\sum_{i_1<i_2<i_3}\operatorname{tr}^{\rm f}(i_1,i_2,i_3) F_4^{\operatorname{tr}(\phi^2)}(i_1,i_2,i_3|1,2,3,4) \\
     &+\lambda_4\sum_{\sigma\in S_{4}/(\mathbb{Z}_4\times S_2)}\operatorname{tr}^{\rm f}(\sigma(1,2,3,4)) F_4^{\operatorname{tr}(\phi^2)}(\sigma(1,2,3,4)|1,2,3,4) \nonumber \,,
\end{align}
where the last sum is to sum over permutations of four particles module the reflection and the cyclic permutation. The explicit calculation show that \eqref{eq:4pttrF22} is indeed correct if $\lambda_2=1,\lambda_3=\lambda_4=2$. 

It is helpful to list explicitly some examples of the Feynman diagrams of the double-ordered blocks appearing in \eqref{eq:4pttrF22}
\begin{equation}\label{eq:4ptYMSFFblocks}
\begin{aligned}
     \includegraphics[width=0.85\linewidth]{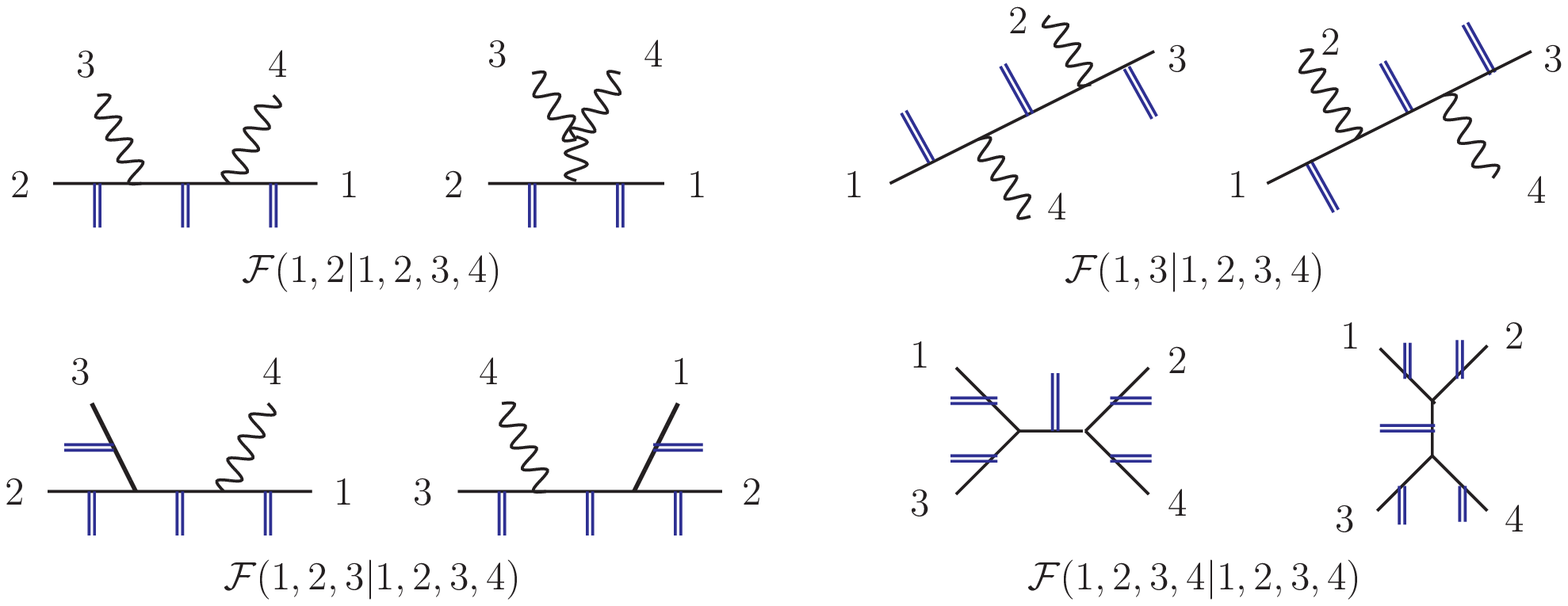}
\end{aligned}
\end{equation}
where we introduce the convention employed in this paper for Feynman diagrams: thin black straight lines for massless scalar, wiggle lines for gluons and double blue lines for $q$; also, drawing multiple double blue $q$-legs in a single diagram means an implicit summation.
From these concrete four-point examples, we observe that the Feynman diagrams of double ordered form factors $F_n^{\operatorname{tr}(\phi^2)}(\alpha|1, \ldots, n)$ can be obtained from the double ordered amplitudes (in the Yang-Mills-scalar theory) $A^{  \rm YMS}(\alpha|1, \ldots, n)$ by inserting the $q$-leg to appropriate positions. 

A more delicate way to organize \eqref{eq:4pttrF22} is as follows 
\begin{equation}\label{eq:4pttrF23}
    F_4^{\operatorname{tr}(F^2)}(1,2,3,4)=\sum_{r=2}^{4}\sum_{i_1<\cdots<i_r}\sum_{\sigma\in S_{r}/\mathbb{Z}_r}\operatorname{tr}^{\rm f}(\sigma(i_1, \ldots, i_r)) F_4^{\operatorname{tr}(\phi^2)}(\sigma(i_1, \ldots, i_r)|1,2,3,4)\,,
\end{equation}
in which we have used the property 
\begin{equation}
    \operatorname{tr}^{\rm f}(\alpha)=(-1)^{|\alpha|} \operatorname{tr}^{\rm f}(\alpha^{-1}),\quad  F^{\operatorname{tr}(\phi^2)}(\alpha|\beta)=(-1)^{|\alpha|}F^{\operatorname{tr}(\phi^2)}(\alpha^{-1}|\beta)\,. 
\end{equation}
It is interesting to notice that a ``flavor-kinematics" duality exists in the sense that when inspecting the last sum in \eqref{eq:4pttrF23} in which $i_1\cdots i_r$ are scalars, the ``closed-trace" of polarizations can be obtained by a simple substitution of the flavour trace  
\begin{equation}
\begin{aligned}
    &\sum_{\sigma\in S_{r}/\mathbb{Z}_r}\operatorname{tr}^{\rm f}(\sigma(i_1, \ldots, i_r)) F_4^{\operatorname{tr}(\phi^2)}(\sigma(i_1, \ldots, i_r)|1,2,3,4) \\
    =& \sum_{\sigma\in S_{r}/\mathbb{Z}_r}\operatorname{tr}_{\rm FL}(\sigma(i_1, \ldots, i_r))\big|_{T^{I_i}\rightarrow \mathrm{f}^{i}} F_4^{\operatorname{tr}(\phi^2)}(\sigma(i_1, \ldots, i_r)|1,2,3,4)\\
    =&\sum_{\sigma\in S_{r}/\mathbb{Z}_r} \big(\mathbf{F}_4^{\operatorname{tr}(\phi^2)}(1,2,3,4) \text{ where } i_1, \ldots, i_r \text{ are scalars}\big)\big|_{T^{I_i}\rightarrow \mathrm{f}^{i}}
\end{aligned}
\end{equation}
where in the last line $\mathbf{F}_4^{\operatorname{tr}(\phi^2)}$ is the color-ordered form factor rather than the double-ordered one. As a result, we can to reorganize the expansion according to the scalar ``skeleton" diagrams, which can be defined by eliminating external gluons in cubic diagrams. For the four-point case mentioned above, one gets 
\begin{equation}
\begin{aligned}
    \mathbf{F}_4^{\operatorname{tr}(\phi^2)}(1^{\phi},2^{\phi},3^{\phi},4^{g})\big|_{T^{I_i}\rightarrow \mathrm{f}^{i}}&\hskip -5pt =f^{123}\big|_{T^{I_i}\rightarrow \mathrm{f}^{i}} {F}_{4}^{\operatorname{tr}(\phi^2)}\Big(\begin{aligned}
      \includegraphics[height=0.075\linewidth]{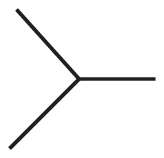}
    \end{aligned}\hskip -3pt \Big)=2 \operatorname{tr}^{\rm f}(1,2,3){F}_{4}^{\operatorname{tr}(\phi^2)}
    \Big(\begin{aligned}
      \includegraphics[height=0.075\linewidth]{fig/cubic3pt.eps}
    \end{aligned}\hskip -3pt \Big)
\end{aligned}
\end{equation}
where the scalar ``skeleton" is nothing but a three point vertex, 
and 
\begin{equation}
\begin{aligned}
    \mathbf{F}_4^{\operatorname{tr}(\phi^2)}&(1^{\phi},2^{\phi},3^{\phi},4^{\phi})\big|_{T^{I_i}\rightarrow \mathrm{f}^{i}}\\
     = & f^{12 \rm x}f^{\rm x 34} \big|_{T^{I_i}\rightarrow \mathrm{f}^{i}} 
    F_4^{\operatorname{tr}(\phi^2)}\Big(
    \begin{aligned}
      \includegraphics[height=0.072\linewidth]{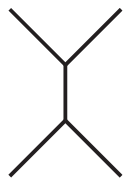}
    \end{aligned}
    \Big) + f^{41 \rm x}f^{\rm x 23} \big|_{T^{I_i}\rightarrow \mathrm{f}^{i}} F_4^{\operatorname{tr}(\phi^2)}\Big(
    \begin{aligned}
      \includegraphics[width=0.078\linewidth]{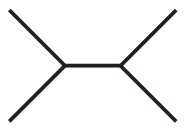}
    \end{aligned}
    \Big)\\
     = & 2\operatorname{tr}^{\rm f}(1,[2,3],4) F_4^{\operatorname{tr}(\phi^2)}\Big(
    \begin{aligned}
      \includegraphics[width=0.078\linewidth]{fig/s4pt.eps}
    \end{aligned}
    \Big) +  2\operatorname{tr}^{\rm f}(1,2,[3,4])F_4^{\operatorname{tr}(\phi^2)}\Big(
    \begin{aligned}
      \includegraphics[height=0.072\linewidth]{fig/t4pt.eps}
    \end{aligned}
    \Big)\,,
\end{aligned}
\end{equation}
where $[i,j]$ means the commutator.
And the explicit Feynman diagrams for each ``skeleton" form are given by
\begin{equation}
\begin{aligned}
    &{F}_{4}^{\operatorname{tr}(\phi^2)}
    \Big(\begin{aligned}
      \includegraphics[height=0.075\linewidth]{fig/cubic3pt.eps}
    \end{aligned}\Big)=\begin{aligned}
      \includegraphics[height=0.085\linewidth]{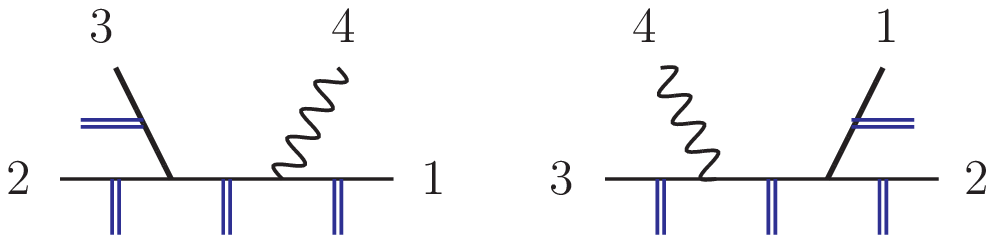}
    \end{aligned} ,\\
    &{F}_{4}^{\operatorname{tr}(\phi^2)}
    \Big(\begin{aligned}
      \includegraphics[width=0.078\linewidth]{fig/s4pt.eps}
    \end{aligned}\Big)=\hskip -8pt\begin{aligned}
      \includegraphics[width=0.16\linewidth]{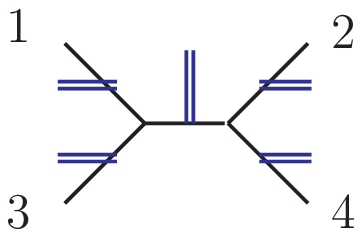}
    \end{aligned}, \quad {F}_{4}^{\operatorname{tr}(\phi^2)}
    \Big(\begin{aligned}
      \includegraphics[height=0.072\linewidth]{fig/t4pt.eps}
    \end{aligned}\Big)=\hskip -10pt \begin{aligned}
      \includegraphics[height=0.11\linewidth]{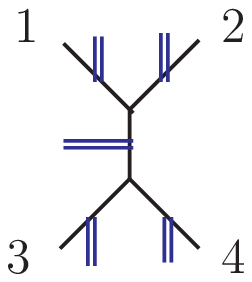}
    \end{aligned}.
\end{aligned}
\end{equation}
Such a ``skeleton" diagram expansion can be generalized to higher points and also has a close relation to color-kinematics duality and double copy, as will be discussed in other works. Below we continue to focus on the decomposition similar to \eqref{eq:4pttrF23}.

\subsection{The general $n$-point case}\label{ssec:trF2proof}
Following \eqref{eq:4pttrF23}, the $n$-point generalization is quite obvious\footnote{Alternatively, the last sum can be replace by $\sum_{\sigma\in S_{t}/(\mathbb{Z}_t\times \mathbb{Z}_2)}\operatorname{tr}^{\rm f}(\sigma(i_1,\ldots,i_t))(1+\delta_{2 t})$ $ F^{\operatorname{tr}(\phi^2)}(\sigma(i_1,\ldots,i_t)|1,\ldots,n)$. The reason why the two particle part is special is that the flavor factor related to $F^{\operatorname{tr}(\phi^2)}(a,b|1 \ldots n)$ is $\delta^{ab}=\operatorname{tr}(T^{I_a}T^{I_b})$, not a product of structure constants.}
\begin{equation}\label{eq:npttrF2}
    F_n^{\operatorname{tr}(F^2)}(1,\ldots,n)=\sum_{r=2}^{n}\sum_{i_1<\cdots<i_r}\sum_{\sigma\in S_{t}/\mathbb{Z}_t}\operatorname{tr}^{\rm f}(\sigma(i_1, \ldots, i_r)) F_n^{\operatorname{tr}(\phi^2)}(\sigma(i_1, \ldots, i_r)|1, \ldots, n)\,.
\end{equation}
And it is straight forward to define a transmuted operator as the ``closed-trace" derivative $\mathcal{D}_{\sigma(i_1,i_2,\ldots,i_r)}$
\begin{equation}
    \mathcal{D}_{(i_1,i_2,\ldots,i_r)}:= \partial_{({\color{red}\bm \epsilon_{i_{1}}}\cdot p_{i_2})}\partial_{(\epsilon_{i_2}\cdot p_{i_3})}\cdots \partial_{(\epsilon_{i_r}\cdot {\color{red}\bm p_{i_{1}}})}\,,
\end{equation}
so that 
\begin{equation}
\begin{aligned}
    &\mathcal{D}_{(i_1, \ldots, i_r)} \operatorname{tr}^{\rm f}(j_1, \ldots, j_r)=1 \Leftrightarrow \exists \kappa \in \mathbb{Z}_{r} \text{ so that } \kappa(i_1, \ldots, i_r)=(j_1, \ldots, j_r)\,,\\
    &\mathcal{D}_{(i_1,\ldots, i_r)} \operatorname{tr}^{\rm f}(j_1,\ldots, j_m)=0 \text{ for any other situations. }
\end{aligned}
\end{equation}
Parallel to \eqref{eq:YMderivative}, we also have\footnote{We can formally distinguish a ``closed-trace" derivative on $\alpha$ and $\alpha^{\scriptscriptstyle \rm T}$, since $\mathcal{D}_{\alpha}=(-1)^{|\alpha|}\mathcal{D}_{\alpha}^{\scriptscriptstyle \rm   T}$. Combining both $\mathcal{D}_{\alpha}$ and $\mathcal{D}_{\alpha}^{\scriptscriptstyle \rm   T}$ together (without distinguishing them) only gives the $(1+\delta_{2t})$ factor in the last footnote.}
\begin{equation}
    \mathcal{D}_{(i_1, \ldots, i_r)}F_n^{\operatorname{tr}(F^2)}(1, \ldots, n)=F_n^{\operatorname{tr}(\phi^2)}(i_1, \ldots, i_r|1, \ldots, n)\,.
\end{equation}

Then we give a proof of such a decomposition formula by examining the factorization properties. 
Within \eqref{eq:npttrF2}, two kind of building blocks are  $F_n^{\operatorname{tr}(\phi^2)}(i_1, \ldots, i_r|1, \ldots, n)$ and $\operatorname{tr}(i_1, \ldots, i_r)$, and we give their factorization properties below. 

First, we focus on the form factors. 
For convenience, we can consider the residue on the $s_{1,\ldots, m}$ pole, on which the double ordered form factor factorize as a double ordered amplitudes and a lower point form factor:
\begin{equation}\label{eq:DOFFfact1}
\begin{aligned}
    \text{Res}[F_n^{\operatorname{tr}(\phi^2)}&(i_1, \ldots, i_r|1, \ldots, n)]_{s_{1,\cdots, m}= 0}\\
    &=A^{ \rm YMS}(i_{1}, \ldots, i_{r'}, I^{+}|1, \ldots, m,I^{+})F_{n{-}m{+}1}^{\operatorname{tr}(\phi^2)}(I^{-},i_{r^{\prime}+1}, \ldots, i_r|I^{-},m{+}1, \ldots, n)\,,
\end{aligned}
\end{equation}
here we have assumed that the interchanged particle is a scalar denoted by $I^+,I^-$. If it is a gluon(denoted by $I^{+}(\epsilon),I^{-}(\bar{\epsilon})$), things will be quite simpler, as the amplitude must be composed of continuous gluons and the scalar skeleton is not changed by the factorization\footnote{The reason why the scalar structure is preserved is that we take only the single trace flavor structure $\operatorname{tr}_{\rm FL}(Y^{i_1}\ldots Y^{i_r})$ so that the scalar "skeleton" must be connected. Thus, if the interchanged particle is a gluon, all the scalars must be in only one side of the factorization.}
\begin{equation}\label{eq:DOFFfact2}
\begin{aligned}
    \text{Res}[F_n^{\operatorname{tr}(\phi^2)}(i_1,&\ldots,i_r|1,\ldots,n)]_{s_{1,\ldots, m}= 0}\\
    &=\sum_{\rm \epsilon\in \epsilon_{\pm}}A^{  \rm YM}(1,\ldots,m,I^{+}(\epsilon))F_{n{-}m{+}1}^{\operatorname{tr}(\phi^2)}(i_1,\ldots,i_r|I^{-}(\Bar{\epsilon}),\ldots,n)\,.
\end{aligned}
\end{equation}

Second,  a ``factorization" related to $\operatorname{tr}(i_1 \ldots i_r)$ is crucial. In particular, we ``factorize" the ``closed-trace" derivatives $\mathcal{D}_{(i_1,\ldots,i_r)}$ into an ``open-trace" derivative and a ``closed-trace" derivatives acting respectively on two blocks
\begin{equation}\label{eq:splitderivative}
    \mathcal{D}_{(i_1,\ldots,i_r)}\cong \widetilde{\mathcal{D}}_{(i_1,\ldots,i_{r^{\prime}},I^{+}(\epsilon))}{\scriptstyle \circ} {\mathcal{D}}_{(I^{-}(\bar{\epsilon}),i_{r^{\prime}+1},\ldots,i_r)}\,,
\end{equation}
the ${\scriptstyle \circ}$ implies a summation over helicity and the $\cong$ symbol means that the two sides of \eqref{eq:splitderivative} coincides when acting on amplitudes, as will be explained in the Appendix~\ref{app:proof dec trF2}. 

To show the consistency of \eqref{eq:npttrF2}, we consider a commutator between taking derivatives and taking residues on physical poles. Such a technique was originally proposed in \cite{Cheung:2017ems}. 
More precisely, we show the following two approaches are equivalent: 
\begin{enumerate}
    \item Taking derivatives before residues
    \begin{equation}\label{eq:cutderivative1}
       \text{Expression 1: } \text{Res}\left[\mathcal{D}_{\sigma(i_1,i_2,\ldots,i_r)}F_n^{\operatorname{tr}(F^2)}(1,\ldots,n)\right]_{s_{1,\ldots,m}= 0}
    \end{equation}
    \item Taking residues  before  derivatives
    \begin{align}\label{eq:cutderivative2}
        \text{Expression 2: }&\mathcal{D}_{\sigma(i_1,i_2,\ldots,i_r)}\left(\text{Res}\big[F_n^{\operatorname{tr}(F^2)}(1,\ldots,n)\big]_{s_{1,\ldots,m}= 0}\right)\\
        =&\mathcal{D}_{\sigma(i_1,i_2,\ldots,i_r)}\left(A^{ \rm YM}(1,\ldots,m,I^{+}(\epsilon))\circ F_{n{-}m{+}1}^{\operatorname{tr}(F^2)}(I^{-}(\bar{\epsilon}),m{+}1,\ldots,n)\right)\,. \nonumber
    \end{align}
\end{enumerate}
Since the form factor is just a rational function of Lorentz products of polarization vectors and momenta, its reasonable to expect that Expression 1 equals to Expression 2. Now we explain why such an identity holds if we have the decomposition \eqref{eq:npttrF2}.

We start from Expression 1, plugging in \eqref{eq:npttrF2} and performing a direct calculation
\begin{equation}
    \text{Res}\left[\mathcal{D}_{\sigma(i_1,i_2,\ldots,i_r)}F_n^{\operatorname{tr}(F^2)}(1,\ldots,n)\right]_{s_{1,\ldots,m}= 0}= \text{Res}\left[F_n^{\operatorname{tr}(\phi^2)}(\sigma(i_1,\ldots,i_r)|1,\ldots,n)\right]_{s_{1,\ldots,m}=0}\,,
\end{equation}
giving 
\begin{equation}\label{eq:cutderivative12}
     A^{  \rm YMS}(i_{1},\ldots,i_{r'},I^{+}|1,\ldots,m,I^{+})F_{n{-}m{+}1}^{\operatorname{tr}(\phi^2)}(I^{-},i_{r^{\prime}+1},\ldots,i_r|I^{-},m{+}1,\ldots,n)\,.
\end{equation}
where we have used the factorization of double ordered form factors \eqref{eq:DOFFfact1} for simplicity. There is no obvious difference for considering the exchanging gluon case ($I$ is a gluon).

Next we try to calculate Expression 2.
Given \eqref{eq:splitderivative}, the next step is trivial: one can plug in the decomposition formula for both form factors and amplitudes, so that the Expression 2 in \eqref{eq:cutderivative2} becomes 
\begin{equation}
    \left(\widetilde{\mathcal{D}}_{(i_1,\ldots,i_r)}A^{  \rm YM}(1,\ldots,m,I^{+}(\epsilon))\right)\left({\mathcal{D}}_{(I^{+}(\epsilon),i_{r^{\prime}+1},\ldots,i_r)}F_{n{-}m{+}1}^{\operatorname{tr}(F^2)}(I^{-}(\bar{\epsilon}),m{+}1,\ldots,n)\right)\,.
\end{equation}
Plugging in the (lower point) decomposition formula for both the amplitude and the form factor, one finally reach 
\begin{align}\label{eq:cutderivative13}
   &\widetilde{\mathcal{D}}_{(i_1,\ldots,i_{r'})}A^{  \rm YM}(1,\ldots,m,I^{+}(\epsilon))= A^{  \rm YMS}(i_{1},\ldots,i_{r'},I^{+}|1,\ldots,m,I^{+}) \\
   \nonumber &{\mathcal{D}}_{(I^{+}(\epsilon),i_{r^{\prime}+1},\ldots,i_r)}F_{n{-}m{+}1}^{\operatorname{tr}(F^2)}(I^{-}(\bar{\epsilon}),m{+}1,\ldots,n)= F_{n{-}m{+}1}^{\operatorname{tr}(\phi^2)}(I^{-},i_{r^{\prime}+1},\ldots,i_r|I^{-},m{+}1,\ldots,n)\,,
\end{align}
of which the product is exactly \eqref{eq:cutderivative12}. 

To summarize, we have shown that assuming the decomposition \eqref{eq:npttrF2} is correct, the Expression 1 and 2 in \eqref{eq:cutderivative1} and \eqref{eq:cutderivative2} are equivalent, which is the most important consistency check for \eqref{eq:npttrF2}. Furthermore, by induction, one can deduce the higher-point decomposition formula from lower-point cases, at least up to special contributions vanishing under all the derivatives $\mathcal{D}_{(i_1,\ldots,i_r)}$. However, gauge invariance and momentum power counting forbid such contributions.\footnote{It is sufficient to consider only the factorizations because there is no $n$-particle gauge invariant contact terms with the correct mass dimension.} This completes the proof for \eqref{eq:npttrF2}, which bases on a similar decomposition for Yang-Mills amplitudes \eqref{eq:YMdecomp} and unitarity \emph{i.e.} factorizations of form factors.

\section{Expanding ${\rm tr}(\phi^2)$ form factors into YMS amplitudes}\label{sec:phi2decomp}
In this section, we present a general expansion of ${\rm tr}(\phi^2)$ form factors with $r$ scalars and $n{-}r$ gluons into YMS amplitudes with $r{+}1$ scalars (including $q$). As we will show, the pure scalar case ($n{=}r$) plays an important role and turns out to be the most difficult step in establishing the expansion for general case, and the inclusion of gluons becomes relatively simple once this is done.

\subsection{Expansion for pure scalar cases}\label{ssec:scalarskeleton}
Let us begin by considering pure-scalar, $r$-point ${\rm tr}(\phi^2)$ form factor, which we denote as $F_r^{{\rm tr}(\phi^2)}(\alpha|\beta)$ with two orderings $\alpha, \beta$. 
We will present an algorithm for its expansion into a linear combination of $(r{+}1)$-point bi-adjoint $\phi^3$ amplitudes. 
The existence of such an expansion per se is not surprising since we can always express trivalent scalar diagrams as linear combinations of bi-adjoint $\phi^3$ amplitudes. However, the key of our construction is to maintain both orderings for the $(r{+}1)$-point amplitude, where the $q$ leg is inserted in $\alpha$ and $\beta$ orderings. 
\subsubsection{The $\alpha=\beta$ cases}
The special case $F_r^{{\rm tr}(\phi^2)}(\alpha|\alpha)$,  say $\alpha=(1,2, \ldots, r)$, turns out to be particularly simple but illuminating. We find that 
\begin{equation} \label{eq pure scalar}
\begin{split}
    &F_r^{{\rm tr}(\phi^2)}(1,2,\ldots,r| 1,2,\ldots,r) \\
    =&\sum_{1\leq a \leq b <r } A(1,2,\ldots,a,q,a{+}1,\ldots,r| 1,2,\ldots,b,q,b{+}1,\ldots,r)\,,
\end{split}
\end{equation}
here and throughout we use $A(\alpha|\beta)$ to represent a double ordered YMS($\phi^3$) amplitudes. 
Note that the RHS of \eqref{eq pure scalar} has $1+2+\cdots+(r{-}1)=r(r{-}1)/2$ terms and the orderings of the amplitudes are all obtained by inserting $q$ into the standard ordering $1,2,\ldots,r$. And let us spell out the simplest examples for $r=3,4$:
\begin{equation} \label{eq pure scalar 3pt}
\begin{split}
    F_3^{{\rm tr}(\phi^2)}(1,2,3| 1,2,3) = & A(1,q,2,3| 1,q,2,3)+A(1,q,2,3| 1,2,q,3)\\
    &+ A(1,2,q,3| 1,2,q,3)\,.
\end{split}
\end{equation}
and
\begin{align} \label{eq pure scalar 4pt}
    &F_4^{{\rm tr}(\phi^2)}(1,2,3,4| 1,2,3,4) = \\
    & A(1,q,2,3,4| 1,q,2,3,4)+A(1,q,2,3,4| 1,2,q,3,4)+A(1,q,2,3,4| 1,2,3,q,4) \nonumber \\
    &+ A(1,2,q,3,4| 1,2,q,3,4)+A(1,2,q,3,4| 1,2,3,q,4) \nonumber  \\
    &+ A(1,2,3,q,4| 1,2,3,q,4). \nonumber
\end{align}

By definition, the LHS of \eqref{eq pure scalar} satisfies the following cyclic and reflection symmetries, which for consistency the RHS should also do:
\begin{enumerate}
    \item Cyclicity: the form factors are defined to satisfy
    $$F_r^{{\rm tr}(\phi^2)}(1,2,\ldots,r| 1,2,\ldots,r)=F_r^{{\rm tr}(\phi^2)}(r,1,2,\ldots,r{-}1| r,1,2,\ldots,r{-}1);$$ On the other hand, the U(1) decoupling relation for those double ordered amplitudes ensures that the RHS also has the cyclicity property.
    \item Reflection:
    we have the reflection symmetry $$F_r^{{\rm tr}(\phi^2)}(1,2,\ldots,r| 1,2,\ldots,r)= (-1)^r F_r^{{\rm tr}(\phi^2)}(1,2,\ldots,r| r,r{-}1,\ldots,1).$$ for the ordered form factor; and if we give the following definition (especially the minus sign marked in red) similar to \eqref{eq pure scalar} when dealing with $F(\alpha|\alpha^{-1})$
\begin{equation}\label{eq pure scalar2}
\begin{split}
    &F_r^{{\rm tr}(\phi^2)}(1,2,\ldots,r| r,r{-}1,\ldots,1) \\
    =&\ {\color{red}-}\sum_{1\leq a \leq b <n } A(1,2,\ldots,a,q,a{+}1,\ldots,r| (1,2,\ldots,b,q,b{+}1,\ldots,r)^{-1}), 
\end{split}
\end{equation}
then the RHS of \eqref{eq pure scalar} also gives $(-1)^{r}$ when taking reflections, due to the reflection property of those $(r{+}1)$-point amplitudes.
The minus sign in \eqref{eq pure scalar2} will be useful below. We also mention another reflection symmetry, which is trivial due to U(1) decoupling, as
$$F_r^{{\rm tr}(\phi^2)}(r,r{-}1,\ldots,1| r,r{-}1,\ldots,1)=F_r^{{\rm tr}(\phi^2)}(1,2,\ldots,r| 1,2,\ldots,r)$$
with
\begin{equation}
\begin{split}
    &F_r^{{\rm tr}(\phi^2)}(r,r{-}1,\ldots,1| r,r{-}1,\ldots,1) \\
    =&\sum_{1\leq b \leq a <r } A(r,r{-}1,\ldots,a{+1},q,a,\ldots,1| r,r{-}1,\ldots,b{+1},q,b,\ldots,1)\,.
\end{split}
\end{equation}

\end{enumerate}

Moreover, the special double sum structure is naturally consistent  with factorizations.
Given the cyclic invariance of \eqref{eq pure scalar}, it's suffice to consider the pole $s_{1,\ldots, m}$. Within the expansion \eqref{eq pure scalar}, only those terms from the $a\geq m$ part of the summation contribute, thus one gets
\begin{equation}\label{eq:purescalarproof}
\begin{split}
    &{\rm Res}[F_r^{{\rm tr}(\phi^2)}(1,2,\ldots,r| 1,2,\ldots,r)]_{s_{1,\ldots, m}} \\
    =&\sum_{m \leq a \leq b <r } {\rm Res}[A(1,2,\ldots,a,q,a{+}1,\ldots,r| 1,2,\ldots,b,q,b{+}1,\ldots,r)]_{s_{1,\ldots, m}}\\
    =&\ A(1,\ldots,m,I^{+}|1,\ldots,m,I^{+})\ \times \\
    & \qquad \sum_{m \leq a \leq b <r }A(I^{-},\ldots,a,q,a{+}1,\ldots,r| I^{-},\ldots,b,q,b{+}1,\ldots,r)\\
    =&\ A(1,\ldots,m,I^{+}|1,\ldots,m,I^{+}) F_{r-m{+}1}^{{\rm tr}(\phi^2)}(I^{-},\ldots,r| I^{-},\ldots,r)\,.
\end{split}
\end{equation}
In the end, we comment that the special double sum structure in Eq.\eqref{eq pure scalar} serves as a building block for the expansions of general cases.

\subsubsection{The $\alpha\neq\beta$ cases: examples}
Next we consider the expansion for $\alpha\neq \beta$, focusing on some concrete examples before moving to most general cases.

To begin with, we consider the case where $\alpha$ and $\beta$ only differs by a {\it partial reflection}, {\it i.e.} for some $1<t<r$ we reverse the sequence $I_2:=(t{+}1, \ldots, r)$ in $\beta$ while keeping $I_1:=(1, \ldots, t)$ in $\alpha$. It turns out the result is remarkably simple:
\begin{equation} \label{eq onereflection}
\begin{split}
    &F_r^{{\rm tr}(\phi^2)}(1,2,\ldots,t,r,r{-}1,\ldots,t{+}1| 1,2,\ldots,t,t{+}1,\ldots,r ) \\
    =& -\sum_{t{+}1\leq b \leq a <r } A(I_1,r,r{-}1,\ldots,a{+}1,q,a,\ldots,t{+}1| 
    I_1,
    t{+}1, \ldots, b, q, b{+}1, \ldots, r)  \\
    & +\sum_{1\leq a \leq b \leq t } A(1,2,\ldots,a,q,a{+}1,\ldots,t,I_2^{-1} |1,2,\ldots,b,q,b{+}1,\ldots,t,I_2 )\,.
\end{split} 
\end{equation}
where in the first line we insert $q$ in the sequence $I_2$ in a way similar to \eqref{eq pure scalar}, and in the second one\footnote{There are boundary cases in the second sum: for $a=t$ we define $(a,q,a{+}1) := (t,q,I_2^{-1})$ and similar for $b=t$.} we insert $q$ as if we are dealing with a $(t{+}1)$-point form factor $\widetilde{F}_{t{+}1}^{{\rm tr}(\phi^2)}(1,2,\ldots,t,I_2^{-1}| 1,2,\ldots,t,I_2)$. In particular,  we add a tilde to emphasize that it is an effective $(t{+}1)$-point one.

As an illustrative example, for $r=4$ with a reflection on $I_2=(34)$, we write
\begin{align}\label{eq onereflection 4pta}
   &F_4^{{\rm tr}(\phi^2)}(1,2,4,3 | 1,2,3,4) \nonumber \\
   =& -A(1,2,4,q,3 | 1,2,3,q,4) \\
   &+A(1,q,2,4,3 | 1,q,2,3,4) +A(1,q,2,4,3 | 1,2,q,3,4) +A(1,2,q,4,3 | 1,2,q,3,4). \nonumber
\end{align}
\begin{figure}[htbp!]
  \centering
  \begin{equation*}
  \begin{aligned}
      \begin{aligned}\includegraphics[width=0.15\linewidth]{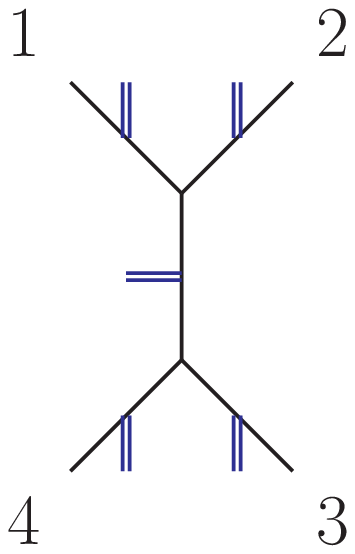}\end{aligned}= \begin{aligned}\includegraphics[width=0.15\linewidth]{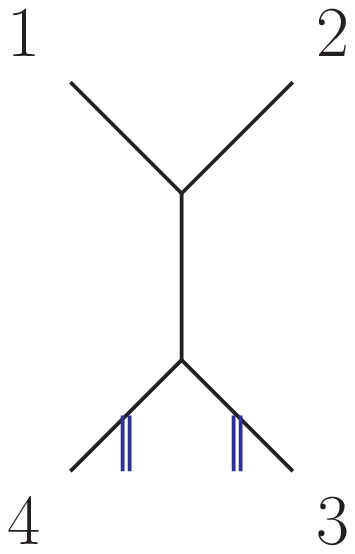}\end{aligned} +\begin{aligned}\includegraphics[width=0.15\linewidth]{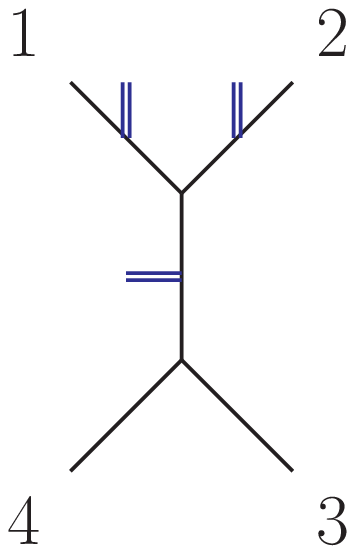}\end{aligned} \left( \begin{aligned}\includegraphics[width=0.15\linewidth]{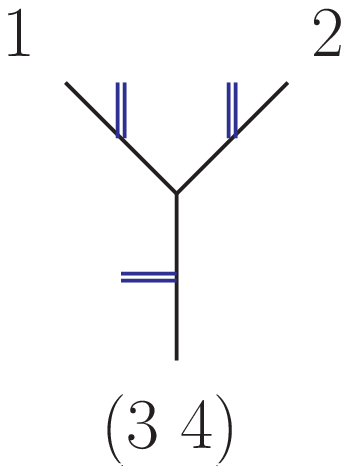}\end{aligned}\right)
  \end{aligned}
  \end{equation*}
  \caption{Feynman diagrams for $F_4^{{\rm tr}(\phi^2)}(1,2,4,3 | 1,2,3,4)$ and its expansion. The term in the large parenthesis is equivalent to the term before it.}\label{fig example onereflection}
\end{figure}
As shown in Figure \ref{fig example onereflection}, the first line of the above expansion computes two graphs where $q$ is inserted on the $3,4$-leg respectively, and the second line gives the effective 3-point form factor $\widetilde{F}_3^{{\rm tr}(\phi^2)}(1,2,(4,3)|1,2,(3,4))$, in which the subgroup $(3,4)$ is regarded as a single particle.
This can be made precise by noticing that $\widetilde{F}_3^{{\rm tr}(\phi^2)}(1,2,(4,3)$ $|1,2,(3,4))\propto s_{3,4}^{-1}$. 
Further more, $\widetilde{F}_3^{{\rm tr}(\phi^2)}(1,2,(4,3)|1,2,(3,4))\propto s_{3,4}^{-1}$ captures all the terms with the $s_{3,4}$ pole in $F_{4}^{{\rm tr}(\phi^2)}(1,2,4,3|1,2,3,4)$:  taking the residue on the $s_{3,4}$ pole shows only the $\widetilde{F}_3^{{\rm tr}(\phi^2)}(1,2,(4,3)|1,2,(3,4))$ part contributes
\begin{align}
    &\text{Res}[F_4^{{\rm tr}(\phi^2)}(1,2,4,3 | 1,2,3,4)]_{s_{3,4}=0} \nonumber \\
    =&\left(A(1,q,2,I^{+}|1,q,2,I^{+}){+}A(1,q,2,I^{+}|1,2,q,I^{+}){+}A(1,2,q,I^{+}|1,2,q,I^{+})\right)\nonumber \\
    &\hskip 9cm \times A(I^{-},4,3|I^{-},3,4)\nonumber \\
    =&\ F_3^{{\rm tr}(\phi^2)}(1,2,I^{+}|1,2,I^{+}) A(I^{-},4,3|I^{-},3,4), \nonumber \\
    = & \text{Res}[\widetilde{F}_3^{{\rm tr}(\phi^2)}(1,2,(4,3)|1,2,(3,4))]_{s_{3,4}=0} 
\end{align}
where the intermediate particles are denoted by $I^+,I^-$ respectively, and the fact that $\text{Res}[A(1,2,4,q,3 | 1,2,3,q,4)]_{s_{3,4}=0}{=}0$ is also used.

Note that there exists an alternative representation 
\begin{align}\label{eq onereflection 4pt}
   &F_4^{{\rm tr}(\phi^2)}(1,2,4,3 | 1,2,3,4) \nonumber \\
   =& A(1,q,2,4,3 | 1,q,2,3,4) \\
   &- A(1,2,q,4,3 | 3,4,q,1,2) -A(1,2,q,4,3 | 3,q,4,1,2)-A(1,2,4,q,3 | 3,q,4,1,2)\,, \nonumber
\end{align}
where the first term on the RHS computes the contribution with $q$ inserted in the subgroup $(1,2)$ and the bottom line is expanding $\widetilde{F}_3^{{\rm tr}(\phi^2)}((1,2),4,3|3,4,(1,2))$. \eqref{eq onereflection 4pta} is equivalent to \eqref{eq onereflection 4pt} so that one can to chose either of them freely, and this will also be the case in the general discussions later. 

\ 

\begin{figure}[t!]
  \centering
  \includegraphics[width=0.9\linewidth]{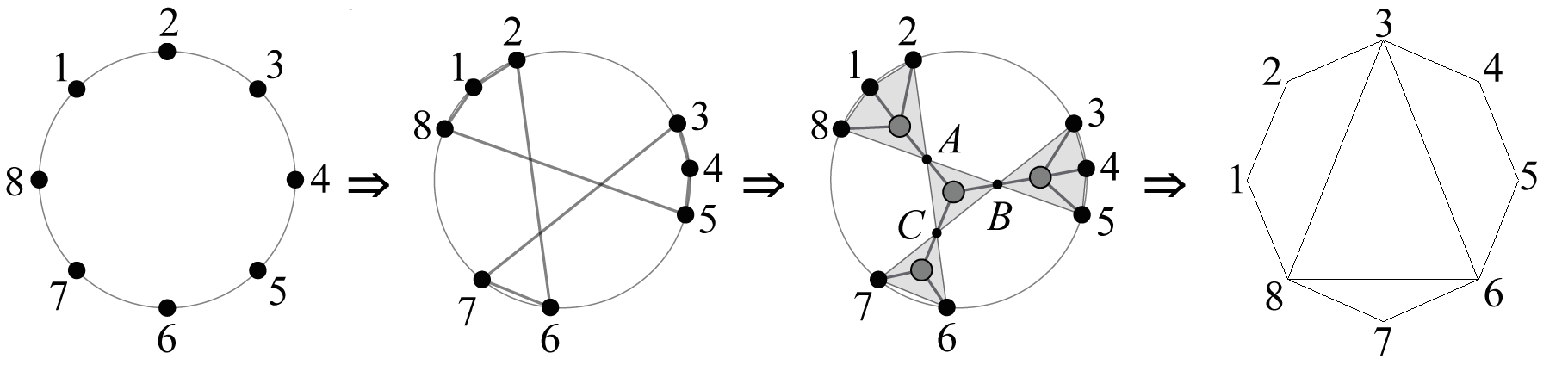}
  \caption{Step by step procedure to draw the mutual partial triangulation for $(1,2,3,4,5,6,7,8| 8,1,2,6,7,3,4,5)$.}\label{1711triangulation}
\end{figure}

\begin{figure}[htbp!]
	\centering
	\subfloat{
	\begin{minipage}[c]{0.4\linewidth}
	\centering
	\begin{tikzpicture}
	\node[regular polygon,
	draw,minimum size=3cm,
	regular polygon sides = 8,rotate=360/16] (p) at (0,0) {};
	\node[above] at (p.side 1) {2}; \node[left] at (p.side 2) {1};
	\node[left] at (p.side 3) {8};  \node[below] at (p.side 4) {7};
	\node[below] at (p.side 5) {6};  \node[right] at (p.side 6) {5};
	\node[right] at (p.side 7) {4}; \node[above] at (p.side 8) {3};
	\draw (p.corner 4) -- (p.corner 1); \draw (p.corner 1) -- (p.corner 6);
	\draw (p.corner 6) -- (p.corner 4);
	\end{tikzpicture}
	\end{minipage}
	}
	\subfloat{
	\begin{minipage}[c]{0.4\linewidth}
	\centering
	\includegraphics[width=0.7\linewidth]{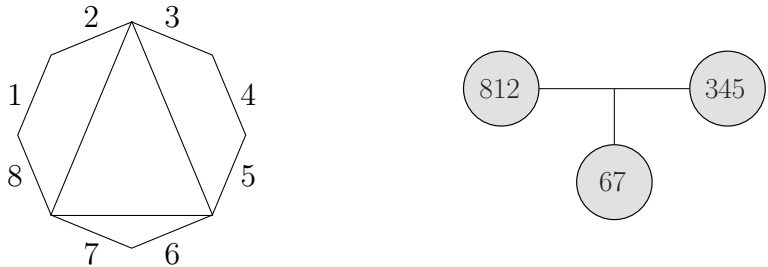}
	\end{minipage}
	}
	\caption{Mutual partial triangulation(left) for $(1,2,3,4,5,6,7,8| 8,1,2,6,7,3,4,5)$ and the corresponding Feynman diagrams(right).} \label{8pt Feynman and tri}
\end{figure}

Now we present another example at 8-point, $F_8^{{\rm tr}(\phi^2)}(8,1,2,6,7,3,4,5 | 1,2,\ldots,8 )$. Importantly, now we introduce an object which captures the main features of Feynman diagrams for $A(\alpha | \beta )$, {\it i.e.} the  {\it mutual partial triangulation} for a given ordered pair $(\alpha|\beta)$\cite{CHY3, Arkani-Hamed:2017mur}. 
We emphasize that the mutual partial triangulation is of particular importance in the discussions below.  
The procedure to draw a mutual partial triangulation is  given by the following steps:
\begin{enumerate}
  \item Draw $n$ points on the boundary of a disk ordered cyclically by $\alpha$.
  \item Draw a closed path of line segments connecting the points in order $\beta$. These line segments can intersect and enclose a set of polygons. 
  \item The interior vertices, as intersections of the aforementioned line segments, 
  correspond to cuts on cubic diagrams.
  \item The cuts are translated to diagonals of the $\alpha$-ordered $n$-gon, forming a mutual partial triangulation.
\end{enumerate}
For our purpose, it is equivalent to draw a mutual partial triangulation for $(\beta|\alpha)$, and a step-by-step procedure to draw such a triangulation for $(1,2,\ldots,8| 8,1,2,6,7,3,4,5)$ is presented in Figure \ref{1711triangulation} where the original picture is found in \cite{Arkani-Hamed:2017mur} and the first three pictures are originally given in \cite{CHY3}. 
Now we further label the edges by the nearest nodes in their anticlockwise direction to represent the corresponding particles in Figure~\ref{8pt Feynman and tri}(left). Note that there are 3 diagonals cutting out 3 parts  $(8,1,2),(3,4,5),(6,7)$ from the polygon, and each diagonal implicates that there is an overall pole in the amplitude, which is  $s_{1,2,8}^{-1},s_{3,4,5}^{-1},s_{6,7}^{-1}$ in this example. Accordingly, we draw the sketch Feynman diagrams for  $A(8,1,2,6,7,3,4,5|1,2,3,4,5,6,7,8)$ in Figure~\ref{8pt Feynman and tri}(right). 

\begin{figure}[htbp!]
	\centering
\begin{equation*}
	\begin{aligned}
		\begin{aligned}
		\includegraphics[width=0.24\linewidth]{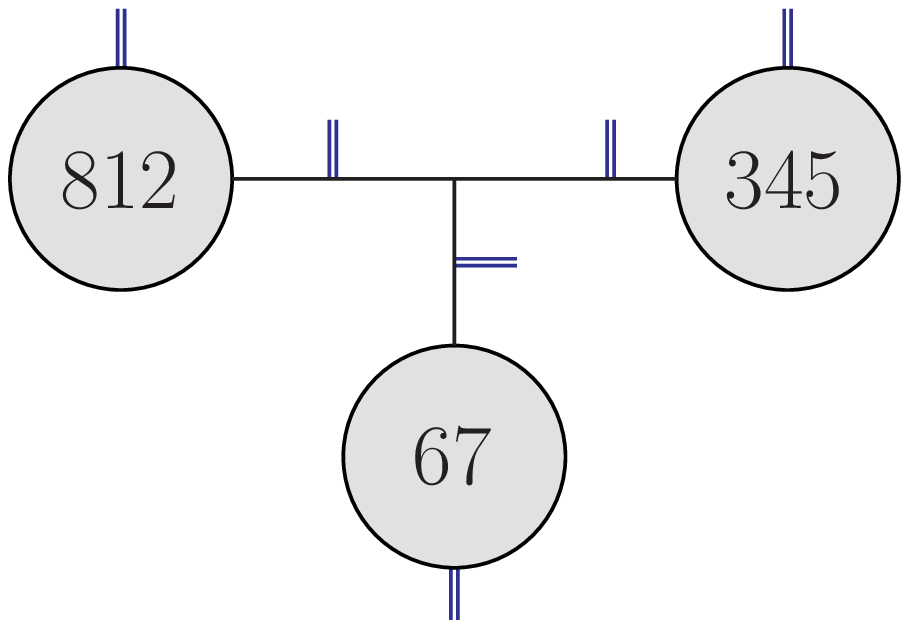}
		\end{aligned} 
		=&\begin{aligned}			\includegraphics[width=0.24\linewidth]{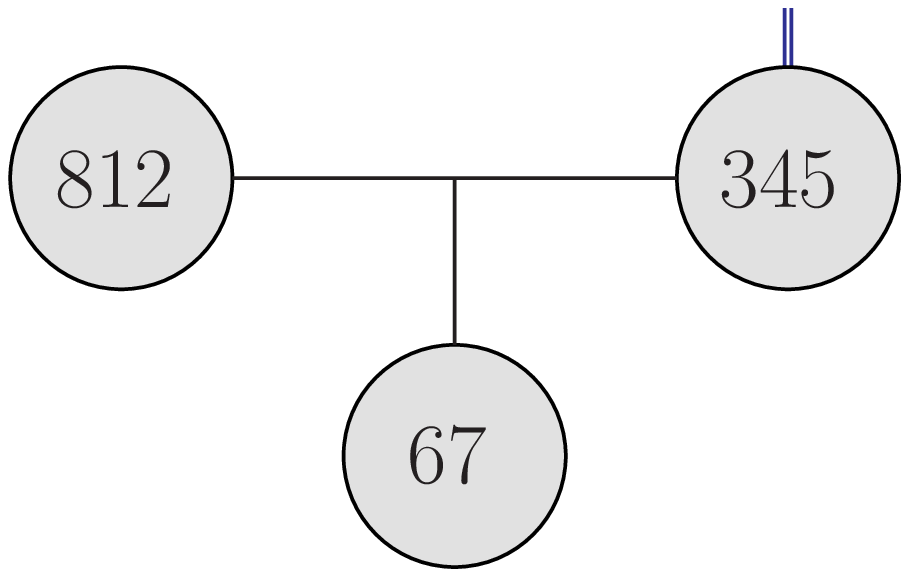}
		\end{aligned}+ \begin{aligned}
			\includegraphics[width=0.24\linewidth]{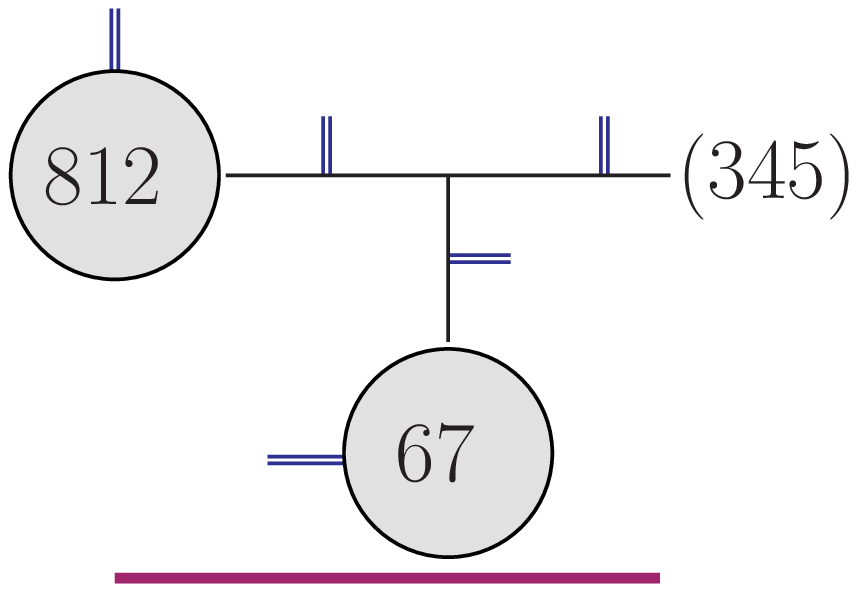}
		\end{aligned}\\
		=& \ldots + {\color[rgb]{0.5,0,0.5}\left(\begin{aligned}
			\includegraphics[width=0.24\linewidth]{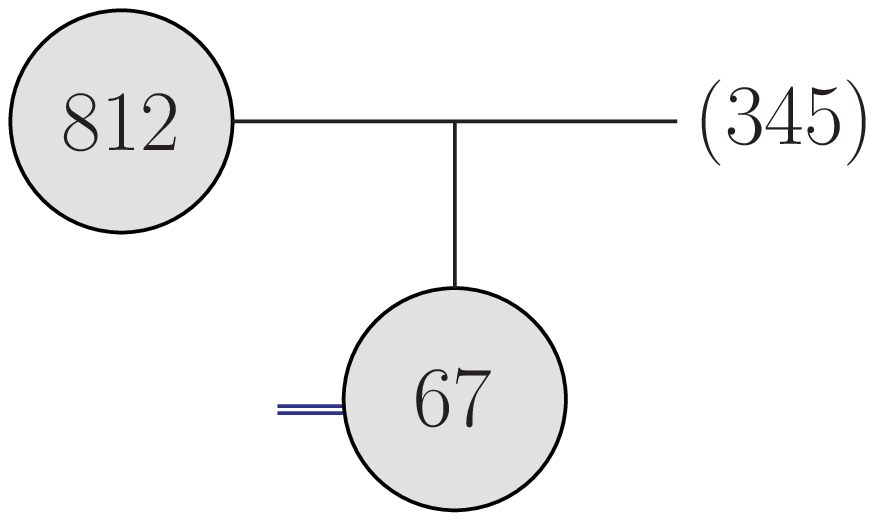}
		\end{aligned}+
	\begin{aligned}
		\includegraphics[width=0.24\linewidth]{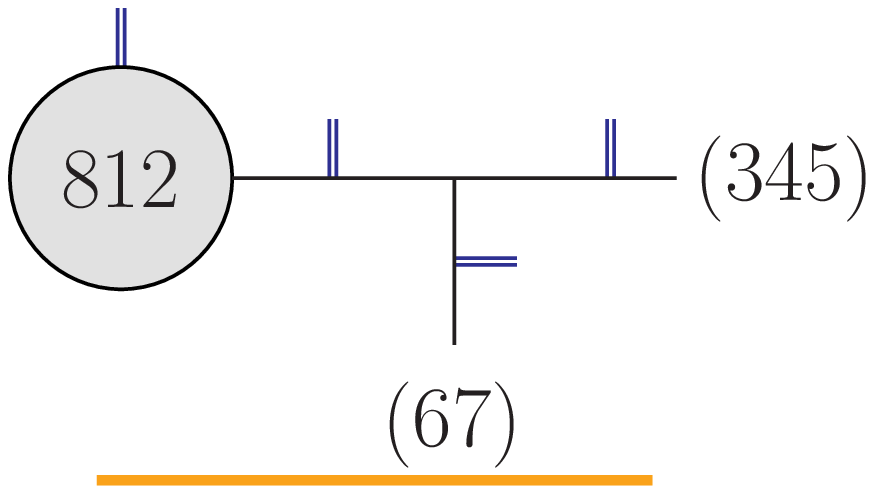}
	\end{aligned} \right)}\\
	=& \ldots\ldots +	{\color[rgb]{1,0.5,0}\left(\begin{aligned}
		\includegraphics[width=0.24\linewidth]{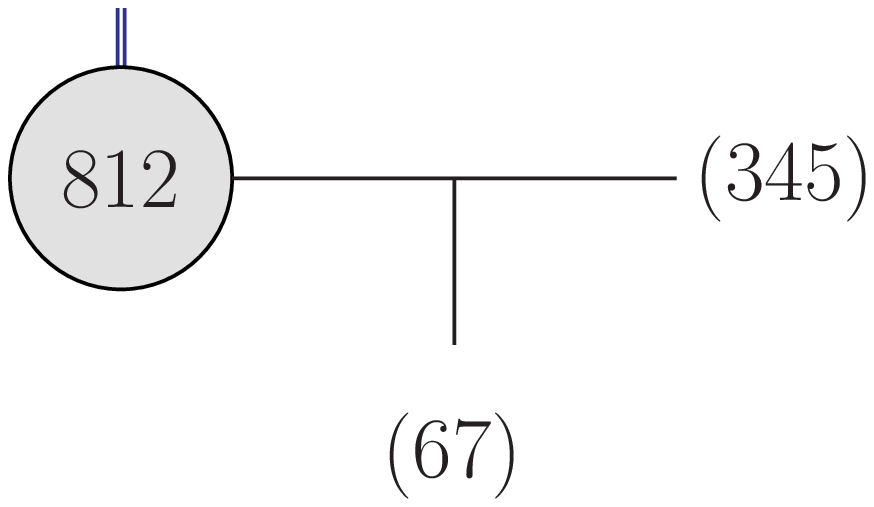}
	\end{aligned} +
	\begin{aligned}
		\includegraphics[width=0.24\linewidth]{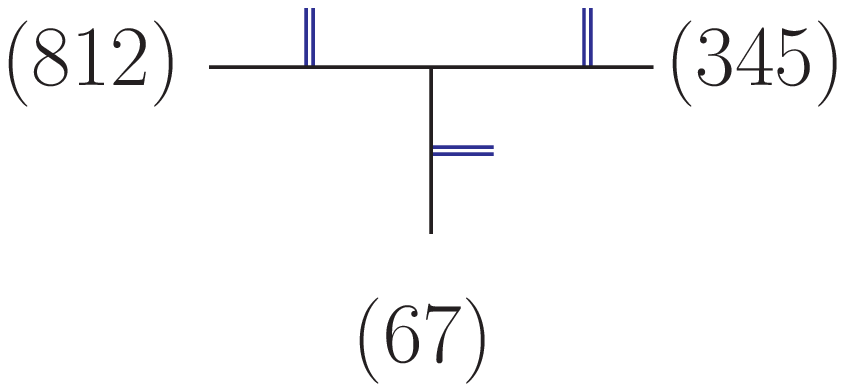}
	\end{aligned} \right)}
	\end{aligned}
\end{equation*}
\centering
\subfloat{
		\begin{minipage}[c]{0.24\linewidth}
			\centering
		\begin{tikzpicture}
		\node[regular polygon,
		draw,minimum size=2.5cm,
		regular polygon sides = 8,rotate=360/16] (p) at (0,0) {};
		\node[above] at (p.side 1) {2}; \node[left] at (p.side 2) {1};
		\node[left] at (p.side 3) {8};  \node[below] at (p.side 4) {7};
		\node[below] at (p.side 5) {6};  
		\node[right] at (p.side 6) {5};
		\node[right] at (p.side 7) {4}; 
		\node[above] at (p.side 8) {3};
		\draw (p.corner 4) -- (p.corner 1); \draw (p.corner 1) -- (p.corner 6);
		\draw (p.corner 6) -- (p.corner 4);  

		\end{tikzpicture}
		\end{minipage}
	}
	\subfloat{
		\begin{minipage}[c]{0.24\linewidth}
			\centering
		\begin{tikzpicture}
		\node[regular polygon,
		draw,minimum size=2.5cm,
		regular polygon sides = 8,rotate=360/16] (p) at (0,0) {};
		\draw[white,line width=2pt] (p.corner 7) -- (p.corner 8); 
		\draw[white,line width=2pt] (p.corner 6) -- (p.corner 7); 
		\draw[white,line width=2pt] (p.corner 1) -- (p.corner 8); 
		\node[above] at (p.side 1) {2}; \node[left] at (p.side 2) {1};
		\node[left] at (p.side 3) {8};  \node[below] at (p.side 4) {7};
		\node[below] at (p.side 5) {6};  
		\draw (p.corner 4) -- (p.corner 1); \draw (p.corner 1) -- (p.corner 6) node[midway, right] {(345)}	;
		\draw (p.corner 6) -- (p.corner 4);  

		\end{tikzpicture}
		\end{minipage}
	}
	\subfloat{
	\begin{minipage}[c]{0.24\linewidth}
		\centering
		\begin{tikzpicture}
			\node[regular polygon,
			draw,minimum size=2.5cm,
			regular polygon sides = 8,rotate=360/16] (p) at (0,0) {};
			\draw[white,line width=2pt] (p.corner 7) -- (p.corner 8); 
			\draw[white,line width=2pt] (p.corner 6) -- (p.corner 7); 
			\draw[white,line width=2pt] (p.corner 1) -- (p.corner 8); 
			\draw[white,line width=2pt] (p.corner 4) -- (p.corner 5);
			\draw[white,line width=2pt] (p.corner 5) -- (p.corner 6);
			\node[above] at (p.side 1) {2}; \node[left] at (p.side 2) {1};
			\node[left] at (p.side 3) {8};  
			\draw (p.corner 4) -- (p.corner 1); \draw (p.corner 1) -- (p.corner 6) node[midway, right] {(345)}	;
			\draw (p.corner 6) -- (p.corner 4) node[midway, below] {(67)};  
			
		\end{tikzpicture}
	\end{minipage}
	}
	\subfloat{
	\begin{minipage}[c]{0.24\linewidth}
		\centering
		\begin{tikzpicture}
			\node[regular polygon,
			draw,minimum size=2.5cm,
			regular polygon sides = 8,rotate=360/16] (p) at (0,4) {};
			\draw[white,line width=2pt] (p.corner 7) -- (p.corner 8); 
			\draw[white,line width=2pt] (p.corner 6) -- (p.corner 7); 
			\draw[white,line width=2pt] (p.corner 1) -- (p.corner 8); 
			\draw[white,line width=2pt] (p.corner 4) -- (p.corner 5);
			\draw[white,line width=2pt] (p.corner 5) -- (p.corner 6);
			\draw[white,line width=2pt] (p.corner 1) -- (p.corner 2);
			\draw[white,line width=2pt] (p.corner 2) -- (p.corner 3);
			\draw[white,line width=2pt] (p.corner 3) -- (p.corner 4);
			\draw (p.corner 4) -- (p.corner 1) node[midway, left] {(812)}; \draw (p.corner 1) -- (p.corner 6) node[midway, right] {(345)}	;
			\draw (p.corner 6) -- (p.corner 4) node[midway, below] {(67)}; 
		\end{tikzpicture}
	\end{minipage}
    }
	\caption{Feynman diagrams computed in each steps of the expansion for $F_8^{{\rm tr}(\phi^2)}(8,1,2,6,7,3,4,5 | 1,2,3,4,5,6,7,8 )$(top), and the reduction of the mutual partial triangulation(bottom). Note that the purple underlined term in the first row becomes the two terms in the purple parenthesis in the second row; and the orange underlined terms in the second row becomes the two terms in the orange parenthesis in the third row. }
	\label{fig:8pt example1}
\end{figure}

Based on the analysis above, we know that there are 4 cubic diagrams for $A(8,1,2,6,7,3,4,5|1,2,3,4,5,6,7,8)$, and the scalar form factor $F(8,1,2,6, 7, 3, 4,5|$ $1,2,3,4,5,6,7,8)$ is obtained by inserting the $q$-leg to all possible scalar lines, leading us to a sum of $4\times 13$ diagrams. 

To write down an expansion of such a form factor, similar to the previous example, we first spell out the contribution from diagrams with $q$ inserted inside the sub-group $(3,4,5)$, so that the rest contributions resemble an effective 6-point form factor $\widetilde{F}_6^{{\rm tr}(\phi^2)}$:
\begin{align}
    &(3,4,5) \text{ contribution } =A(8,1,2,6,7,3,q,4,5 | 1,2,3,q,4,5,6,7,8 ) +  \\
    & A(8,1,2,6,7,3,q,4,5 | 1,2,3,4,q,5,6,7,8 )+A(8,1,2,6,7,3,4,q,5 | 1,2,3,4,q,5,6,7,8 ) \nonumber\\ \nonumber\\
    & F_8^{{\rm tr}(\phi^2)}(8,1,2,6,7,3,4,5 | 1,2,3,4,5,6,7,8 ) - (3,4,5) \text{ contribution }, \nonumber \\
      = &\widetilde{F}_6^{{\rm tr}(\phi^2)}(8,1,2,6,7,(3,4,5) | 1,2,(3,4,5),6,7,8 ),
\end{align}
where the operation of subtracting the $(3,4,5)$ contribution means diagrammatically erasing edges $3,4,5$ in the mutual partial triangulation, see the second polygon in Figure~\ref{fig:8pt example1}(bottom). This gives us a mutual partial triangulation for a 6-point form factor regarding the sub-group $(3,4,5)$ as a single particle.
Importantly, such an effective ``6-point" form factor is proportional to $s_{3,4,5}^{-1}$.
Furthermore, we have the following property \textbf{(without taking residues)}
\begin{equation}\label{eq:f6subgroup}
\begin{aligned}
    \widetilde{F}_6^{{\rm tr}(\phi^2)}&(8,1,2,6,7,(3,4,5) | 1,2,(3,4,5),6,7,8 )\\
    &=\frac{1}{s_{3,4,5}}F_6^{{\rm tr}(\phi^2)}(8,1,2,6,7,I^{+} | 1,2,I^{+},6,7,8)A(3,4,5,I^{-}|3,4,5,I^{-})\,.
\end{aligned}
\end{equation}
To make the expressions in \eqref{eq:f6subgroup} precise, in $A(3,4,5,I^{-}|3,4,5,I^{-})$ we use the momentum conservation to eliminate $I_{-}$ so that $A(3,4,5,I^{-}|3,4,5,I^{-})=1/s_{3,4}+1/s_{4,5}$, and in $F_6^{{\rm tr}(\phi^2)}(8,1,2,6,7,I^{+}| 1,2,I^{+},6,7,8)$ we replace $s_{I^{+}\ldots}$ with $s_{3,4,5,\ldots}$. 

Next, the contribution from diagrams with $q$ inserted inside the subgroups $(1,2,8)$ and $(6,7)$ respectively can be given in a similar way, with an effective 3-point form factor $\widetilde{F}_{3}^{\operatorname{tr}(\phi^2)}((8,1,2),(6,7),(3,4,5) | (3,4,5),(6,7),(8,1,2) )$ remained. Again, it is proportional to $s^{-1}_{3,4,5}s^{-1}_{6,7}s^{-1}_{8,1,2}$, guaranteeing that subgroups $(3,4,5),(6,7),(1,2,8)$ can all be treated as single particles. 
The Feynman diagrams sketching the successive reduction procedure are shown in Figure~\ref{fig:8pt example1}(top).

For the final step of our expansion, one should be careful since it is slightly different from a ``real'' 3-point form factor, the contribution is given by:
\begin{equation}
\begin{split}\label{eq:F3effectivefor8pt}
    \widetilde{F}_{3}^{\operatorname{tr}(\phi^2)}&( (8,1,2),(6,7),(3,4,5) | (3,4,5),(6,7),(8,1,2) )=\\
    &-A( (8,1,2),q,(6,7),(3,4,5) | (3,4,5),(6,7) \shuffle q ,(8,1,2) ) \\
    &-A( (8,1,2),(6,7),q,(3,4,5) | (3,4,5),q,(6,7) ,(8,1,2) ).
\end{split}
\end{equation}
{\it i.e.} $q$ needs to run into $(6,7)$ in the second ordering. The reason is that if we did not do this shuffle, then the RHS of \eqref{eq:F3effectivefor8pt} is no longer proportional to $s_{6,7}^{-1}$. Only the special shuffle product can help us achieve this. 

This is the end of the expansion of the 8-point form factor. 
Also, a step by step reduction of the mutual partial triangulation corresponds to the construction above is given in Figure \ref{fig:8pt example1}(bottom).

\begin{figure}[htbp!]
	\centering
\begin{equation*}
	\begin{aligned}
		\begin{aligned}
		\includegraphics[width=0.24\linewidth]{fig/8pt_example1.eps}
		\end{aligned} 
		=&\begin{aligned}			\includegraphics[width=0.24\linewidth]{fig/8pt_example2.eps}
		\end{aligned}+ \begin{aligned}
			\includegraphics[width=0.24\linewidth]{fig/8pt_example3.eps}
		\end{aligned}\\
		=& \ldots +{\color[rgb]{0.5,0,0.5} \left(\begin{aligned}
			\includegraphics[width=0.24\linewidth]{fig/8pt_example4.eps}
		\end{aligned}+
	\begin{aligned}
		\includegraphics[width=0.24\linewidth]{fig/8pt_example5.eps}
	\end{aligned}\right)} \\
	=& \ldots\ldots +	{\color[rgb]{1,0.5,0}\left(\begin{aligned}
	\includegraphics[width=0.24\linewidth]{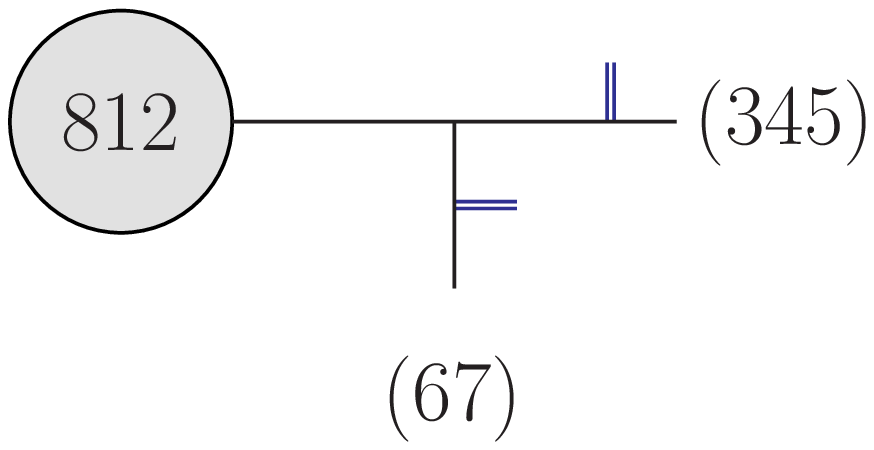}
\end{aligned} +	\begin{aligned}
	\includegraphics[width=0.19\linewidth]{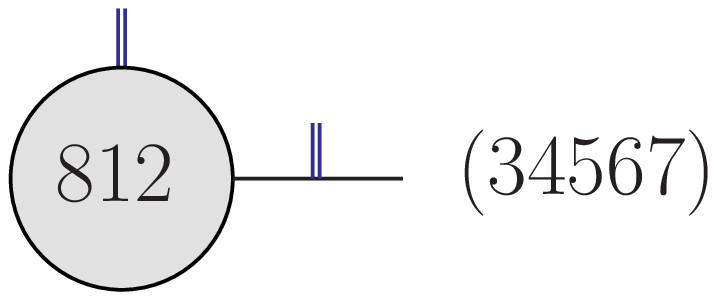}
	\end{aligned}\right)}
	\end{aligned}
\end{equation*}
\centering
\subfloat{
		\begin{minipage}[c]{0.24\linewidth}
			\centering
		\begin{tikzpicture}
		\node[regular polygon,
		draw,minimum size=2.5cm,
		regular polygon sides = 8,rotate=360/16] (p) at (0,0) {};
		\node[above] at (p.side 1) {2}; \node[left] at (p.side 2) {1};
		\node[left] at (p.side 3) {8};  \node[below] at (p.side 4) {7};
		\node[below] at (p.side 5) {6};  
		\node[right] at (p.side 6) {5};
		\node[right] at (p.side 7) {4}; 
		\node[above] at (p.side 8) {3};
		\draw (p.corner 4) -- (p.corner 1); \draw (p.corner 1) -- (p.corner 6);
		\draw (p.corner 6) -- (p.corner 4);  

		\end{tikzpicture}
		\end{minipage}
	}
	\subfloat{
		\begin{minipage}[c]{0.24\linewidth}
			\centering
		\begin{tikzpicture}
		\node[regular polygon,
		draw,minimum size=2.5cm,
		regular polygon sides = 8,rotate=360/16] (p) at (0,0) {};
		\draw[white,line width=2pt] (p.corner 7) -- (p.corner 8); 
		\draw[white,line width=2pt] (p.corner 6) -- (p.corner 7); 
		\draw[white,line width=2pt] (p.corner 1) -- (p.corner 8); 
		\node[above] at (p.side 1) {2}; \node[left] at (p.side 2) {1};
		\node[left] at (p.side 3) {8};  \node[below] at (p.side 4) {7};
		\node[below] at (p.side 5) {6};  
		\draw (p.corner 4) -- (p.corner 1); \draw (p.corner 1) -- (p.corner 6) node[midway, right] {(345)}	;
		\draw (p.corner 6) -- (p.corner 4);  

		\end{tikzpicture}
		\end{minipage}
	}
	\subfloat{
	\begin{minipage}[c]{0.24\linewidth}
		\centering
		\begin{tikzpicture}
			\node[regular polygon,
			draw,minimum size=2.5cm,
			regular polygon sides = 8,rotate=360/16] (p) at (0,0) {};
			\draw[white,line width=2pt] (p.corner 7) -- (p.corner 8); 
			\draw[white,line width=2pt] (p.corner 6) -- (p.corner 7); 
			\draw[white,line width=2pt] (p.corner 1) -- (p.corner 8); 
			\draw[white,line width=2pt] (p.corner 4) -- (p.corner 5);
			\draw[white,line width=2pt] (p.corner 5) -- (p.corner 6);
			\node[above] at (p.side 1) {2}; \node[left] at (p.side 2) {1};
			\node[left] at (p.side 3) {8};  
			\draw (p.corner 4) -- (p.corner 1); \draw (p.corner 1) -- (p.corner 6) node[midway, right] {(345)}	;
			\draw (p.corner 6) -- (p.corner 4) node[midway, below] {(67)};  
			
		\end{tikzpicture}
	\end{minipage}
	}
	\subfloat{
	\begin{minipage}[c]{0.24\linewidth}
		\centering
		\begin{tikzpicture}
			\node[regular polygon,
			draw,minimum size=2.5cm,
			regular polygon sides = 8,rotate=360/16] (q) at (0,0) {};
			\draw[white,line width=2pt] (q.corner 7) -- (q.corner 8); 
			\draw[white,line width=2pt] (q.corner 6) -- (q.corner 7); 
			\draw[white,line width=2pt] (q.corner 1) -- (q.corner 8); 
			\draw[white,line width=2pt] (q.corner 4) -- (q.corner 5);
			\draw[white,line width=2pt] (q.corner 5) -- (q.corner 6);
			\node[above] at (q.side 1) {2}; \node[left] at (q.side 2) {1};
			\node[left] at (q.side 3) {8};  
			\draw (q.corner 4) -- (q.corner 1) node[midway, right] {(34567)};
		\end{tikzpicture}
	\end{minipage}
	}
	\caption{Feynman diagrams computed in each steps of an alternative expansion for $F_8^{{\rm tr}(\phi^2)}(8,1,2,6,7,3,4,5 | 1,2,3,4,5,6,7,8 )$(top), and the reduction of the mutual partial triangulation(bottom).}
	\label{fig:8pt example2}
\end{figure}

In addition, we would like to emphasize that there are several ways of performing the reductions and they are all equivalent in the end. For instance, we have an alternative construction here: compute the contributions from diagrams with $q$ inserted in the subgroups  $(3,4,5),(6,7),((3,4,5),(6,7))$ successively and end up with an effective 4-point form factor $\widetilde{F}_4^{{\rm tr}(\phi^2)}(8,1,2,(6,7,3,4,5)|8,1,2,(3,4,5,6,7))$ reading
{\small 
\begin{align}
    & \widetilde{F}_4^{{\rm tr}(\phi^2)}(8,1,2,(6,7,3,4,5)|8,1,2,(3,4,5,6,7)) {=} A(8,q,1,2,(6,7,3,4,5)|8,q,1,2,(3,4,5,6,7)) \nonumber \\
    &{+}A(8,q,1,2,(6,7,3,4,5)|8,1,q,2,(3,4,5,6,7)){+}A(8,q,1,2,(6,7,3,4,5)|8,1,2,q,(3,4,5,6,7))\nonumber \\ 
    &{+} A(8,1,q,2,(6,7,3,4,5)|8,1,q,2,(3,4,5,6,7)) {+}A(8,1,q,2,(6,7,3,4,5)|8,1,2,q,(3,4,5,6,7))\nonumber \\
    &{+} A(8,1,2,q,(6,7,3,4,5)|8,1,2,q,(3,4,5,6,7))
\end{align}
}
See Figure \ref{fig:8pt example2} for illustrative Feynman diagrams and the mutual partial triangulation.

\subsubsection{The $\alpha \neq \beta$ cases: a general algorithm} \label{ssec:expand scalar general}
We now present the construction process for a pure scalar ${\rm tr}(\phi^2)$ form factor with any given ordered pair $(\alpha| \beta)$. 
Long story short, by inserting the $q$-leg in some subgroups separated by diagonal lines in the partial mutual triangulation, one can effectively drop off these subgroups and turn the high-point problem into a lower-point one. 

As illustrated by the examples given above, the concrete algorithm can be summarized as follows:

\textbf{Step (1)}  Draw the mutual partial triangulation corresponding to $(\alpha|\beta)$.

\begin{figure}[htbp!]
	\centering
	\subfloat[]{
		\begin{minipage}[c]{0.3\linewidth}
			\centering
			\begin{tikzpicture}
				\coordinate (p) at (0,0);
				\draw (p) circle[radius=1.5cm];
				\draw (100:1.5cm) -- (200:1.5cm); \draw (260:1.5cm) -- (200:1.5cm);
				\draw (100:1.5cm) -- (260:1.5cm);
				\draw (30:1.5cm) -- (300:1.5cm); 
				\node[right] at (0:0.1cm) {...};
				\node[right] at (0:1.5cm) {$\alpha_c$};
				\node[right] at (345:1.5cm) {$\alpha_{c+1}$};
				\node[right] at (330:1.5cm) {$\ldots$};
				\node[right] at (315:1.5cm) {$\alpha_{d}$};
				\filldraw (5:1.5cm) circle (.05)
                (355:1.5cm) circle (.05)
                (345:1.5cm) circle (.05)
                (315:1.5cm) circle (.05) (325:1.5cm) circle (.05);
                \node[left] at (195:1.5cm) {$\alpha_1$};
			    \node[left] at (180:1.5cm) {$\alpha_2$};
			    \node[left] at (165:1.5cm) {$\ldots$};
			    \filldraw (200:1.5cm) circle (.05)  (190:1.5cm) circle (.05) (180:1.5cm) circle (.05);
			\end{tikzpicture}
		\end{minipage}
	}
	\subfloat[]{
		\begin{minipage}[c]{0.3\linewidth}
			\centering
			\begin{tikzpicture}
				\coordinate (p) at (0,0);
				\draw (100:1.5cm) -- (200:1.5cm); \draw (260:1.5cm) -- (200:1.5cm);
				\draw (100:1.5cm) -- (260:1.5cm); 
				\node[right] at (0:0.1cm) {...};
				\draw (30:1.5cm) arc(30:300:1.5cm) ; 
				\draw (30:1.5cm) -- (300:1.5cm); 	
			\end{tikzpicture}
		\end{minipage}
	}
	\subfloat[]{
		\begin{minipage}[]{0.3\linewidth}
			\centering
			\begin{tikzpicture}
			\coordinate (p) at (0,0);
			\draw (100:1.5cm) --node[right] {$I_t$} (200:1.5cm) ; 
			\draw (100:1.5cm) arc(100:200:1.5cm) ; 
			\node[left] at (195:1.5cm) {$I_1$};
			\node[left] at (180:1.5cm) {$I_2$};
			\node[left] at (165:1.5cm) {$\ldots$};
			\draw[white] (210:1.5cm) arc(210:360:1.5cm) ; 
			\filldraw (200:1.5cm) circle (.05)  (190:1.5cm) circle (.05) (180:1.5cm) circle (.05);
			\end{tikzpicture}
		\end{minipage}
	}
	\caption{The reduction of the mutual partial triangulation for the general expansion of algorithm. (c) presents a special choice of the final block (covered in Step (4) of the algorithm) to be $\widetilde{F}_r^{{\rm tr}(\phi^2)}(I_1,I_2,\ldots,I_t|I_1^\prime,I_2^\prime,\ldots,I_t^\prime)$ with  $I_1{=}\alpha_1,I_2{=}\alpha_2,\ldots,I_{t{-}1}{=}\alpha_{t{-}1}$ and $I_t{=}(\alpha\backslash (\alpha_1,\alpha_2,\ldots,\alpha_{t{-}1}))$.} \label{fig decom of general cases}
\end{figure}
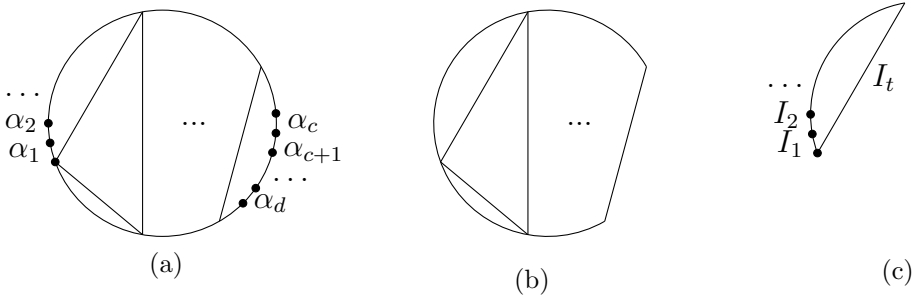

\textbf{Step (2)} In the mutual partial triangulation diagram, consider a set of continuous edges $I_1=(\alpha_c,\alpha_{c+1},\ldots,\alpha_{d})$ cut out by a single diagonal line first, see Figure \ref{fig decom of general cases}(a). There are only two possibilities relevant to such a partial mutual triangulation: either $I_1$ or $I_1^{-1}$ must be a sub-ordering in $\beta$. 
    
To give the contribution from diagrams with $q$ inserted in $I_1$, namely the $I_1$ contribution, we write down the following special double sum for $\phi^3$ amplitudes with $q$ inserted in the sequence $I_1$  similar to \eqref{eq pure scalar} 
{
    \begin{equation*} 
    \begin{aligned}
    &I_1 \text{ contribution}=\\
    &\sum_{c\leq a  <d } A(\alpha_c,\ldots,\alpha_a,q,\alpha_{a{+}1},\ldots,\alpha_d,\ldots| \alpha_c,\ldots,\alpha_{a},(\alpha_{a{+}1},\ldots,\alpha_{d-1})\shuffle q,\alpha_d,\ldots).
    \end{aligned}
\end{equation*}}
    Or if $I_1$ and $I_1^{-1}$ appear in $\alpha$ and $\beta$ respectively, we write 
{
    \begin{equation*} 
    \begin{aligned}
    &I_1 \text{ contribution}=\\
    -&\sum_{c\leq a  <d } A(\alpha_c,\ldots,\alpha_a,q,\alpha_{a{+}1},\ldots,\alpha_d,\ldots| (\alpha_c,\ldots,\alpha_{a},(\alpha_{a{+}1},\ldots,\alpha_{d-1})\shuffle q,\alpha_d)^{-1},\ldots).
    \end{aligned}
\end{equation*}}
In a word, a reversed ordering of the subgroup gives an extra minus sign. 
    
After subtracting the $I_1$ contribution given above, one can now regard the subgroup $(\alpha_c,\alpha_{c+1},\ldots,\alpha_{d})$ as a single particle, which means 
\begin{equation*}
    F_r^{\operatorname{tr}(\phi^2)}(\alpha|\beta)-I_1 \text{ contribution } \propto s_{I_1}^{-1}\,.
\end{equation*}
Such a difference can be defined as an effective lower-point form factor 
$F^{\operatorname{tr}(\phi^2)}_{r^{\prime}}(\alpha|\beta)$ with $r^{\prime}=r{-}|I_1|{+}1$ and $I_1$ treated as a single particle in $\widetilde{F}_{r'}(\alpha|\beta)$. 
On the mutual partial triangulation diagram, this means we now discard the edges $\alpha_c,\ldots,\alpha_d$ and keep only the diagonal cutting them out. Thus, we obtain a lower point mutual partial triangulation diagram with less diagonals for $\widetilde{F}_{r'}(\alpha|\beta)$, see Figure~\ref{fig decom of general cases}(b). 

\textbf{Step (3)} Repeat the Step (2) and reduce the number of diagonals in the mutual partial triangulation, until we reach a polygon with no diagonal lines. This Step is similar to Step (2): we still consider a set of continuous edges in the (reduced) mutual partial triangulation. The difference is that now the edges could represent subgroups, and one should be particular careful about inserting $q$ into the subgroups. 

For example, we deal with the following effective form factor 
$$\widetilde{F}_{s}^{\operatorname{tr}(\phi^2)}(I_1,I_2,I_{3},\kappa|I_1^{\prime},I_2^{\prime},I_{3}^{\prime},\rho)\propto (s_{I_1}s_{I_2}s_{I_3})^{-1},$$
where $I_{i}$ and $I_{i}^{\prime}$ for $i=1,2,3$ are (different) ordered subgroups containing the same set of particles, and $\kappa\neq\rho$ are different sub-orderings of the other particles. In the mutual partial triangulation diagram for $\widetilde{F}_s$, $I_i$ are edges and there exists a diagonal dividing $I_1$, $I_2, I_3$ and the rest of the polygon. Cutting out such a quadrilateral, composed of $I_1,I_2,I_3$ and the diagonal, gives the following contribution
\begin{equation*}
\begin{aligned}
    (I_1,I_2,I_3) \text{ contribution} = & A(I_1,q,I_2,I_{3},\kappa|I_1^{\prime},I_2^{\prime}\shuffle q,I_{3}^{\prime},\rho)\\
    &+ A(I_1,I_2,q,I_{3},\kappa|I_1^{\prime},I_2^{\prime}, q,I_{3}^{\prime},\rho)\,.
\end{aligned}
\end{equation*}
The subtlety here is that the $q$-leg has to be shuffled into subgroups in the right ordering, while $q$ should only be placed between subgroups in the left ordering. The reason why we do this shuffling is to keep the $(I_1,I_2,I_3) \text{ contribution}$ also proportional to $(s_{I_1}s_{I_2}s_{I_3})^{-1}$, as $\widetilde{F}_s$ is.

With now the $(I_1,I_2,I_3) \text{ contribution}$ properly defined, 
we  subtract it from $\widetilde{F}_{s}^{\operatorname{tr}(\phi^2)}$, giving the following $s^{\prime}{=}(s{-}|I_1|{-}|I_2|{-}|I_3|{+}3)$-point effective form factor
\begin{equation*}
\begin{aligned}
    \widetilde{F}^{\operatorname{tr}(\phi^2)}_{s^{\prime}}&((I_1,I_2,I_3),\kappa|(I_1^{\prime},I_2^{\prime},I_3^{\prime}),\rho)\\
    &=\widetilde{F}_{s}^{\operatorname{tr}(\phi^2)}(I_1,I_2,I_{3},\kappa|I_1^{\prime},I_2^{\prime},I_{3}^{\prime},\rho)-(I_1,I_2,I_3) \text{ contribution},
\end{aligned}
\end{equation*}
which is proportional to $s_{I_1 I_2 I_3}^{-1}\times (s_{I_1}s_{I_2}s_{I_3})^{-1}$. 

\textbf{Step (4)}
At last, a ``final" polygon with no diagonal is obtained, of which the edges are denoted by $I_1,I_2,\ldots,I_t$ as in Figure \ref{fig decom of general cases}(c). Then the final contribution is given by
\begin{equation*}
    \sum_{1\leq a  <t } A(I_1,\ldots,I_a,q,I_{a{+}1},\ldots,I_t| I_1^\prime,\ldots,I_a^\prime,(I_{a{+}1}^\prime,\ldots,I_{t{-}1}^\prime) \shuffle q,I_t^\prime),
\end{equation*}
or the reversed version with a minus sign
\begin{equation*}
    -\sum_{1\leq a  <t } A(I_1,\ldots,I_a,q,I_{a{+}1},\ldots,I_t| \big(I_1^\prime,\ldots,I_a^\prime,(I_{a{+}1}^\prime,\ldots,I_{t{-}1}^\prime) \shuffle q,I_t^\prime\big)^{-1})\,.
\end{equation*}
Still, in the second ordering the sequence $I_a^\prime$ contains the same particle as $I_a$ but they can be different up to possible permutations.

\vspace{3pt}

In the end, we emphasize again that there are other expansions, for example the ones ending up with a different polygon in Step (4) in the algorithm, and all possible expansions are equivalent. 
In practice we usually specify one of the constructions as our convention in sec.~\ref{sec:expand and CHY} by requiring the final contribution to  correspond to the last polygon (without any diagonals) contains the last edge $i_r$. 

\subsection{From pure scalar cases to ${\rm tr}(\phi^2)$ form factors with gluons}\label{ssec:withgluons}
Now we move on to the expansion of ${\rm tr}(\phi^2)$ form factors which contain arbitrary number of gluons. As mentioned before, such an expansion is closely related to its scalar skeleton, {\it i.e.} the expansion of a pure-scalar form factor with the same scalar double orderings. We give an ``inserting rule" which transforms the pure-scalar result into one with gluons, thus provide a complete expansion for ${\rm tr}(\phi^2)$ form factors appear on the RHS of \eqref{eq:npttrF2}.

\subsubsection{Rules for inserting gluons}

Let us denote such a general form factor as
$F_n^{{\rm tr}(\phi^2)}(\alpha(i_1,i_2,\ldots,i_r)|i_1,\mathcal{G}_{i_1},i_2,\mathcal{G}_{i_2} \ldots,i_r, \mathcal{G}_{i_r})$ where we label the scalars by $i_1<i_2<\cdots<i_r$ and gluons between scalar pair $i_a,i_{a{+}1}$ as $\mathcal{G}_{i_a}$. Recall the expansion of its scalar skeleton: 
\begin{equation} 
\begin{split}
     &F_r^{{\rm tr}(\phi^2)}(\alpha(i_1,i_2,\ldots,i_r)| i_1,i_2,\ldots,i_r) \\
     =& \sum_{a,b} A(\alpha_{1},\alpha_{2},\ldots,\alpha_{a},q,\alpha_{a{+}1},\ldots,\alpha_{r} |  i_1,i_2,\ldots,i_b,q,i_{b{+}1},\ldots,i_r)\,,
\end{split}
\end{equation}
with certain $a,b$ determined in the last subsection.
We emphasize again that it is crucial to have both orderings for $r$ scalars in each $\phi^3$ amplitude the same as those in the form factor. Given such a skeleton result, we propose that the expansion of the form factor with gluons is given by a surprisingly simple rule: literally place every gluons set into the second ordering and sum over the shuffle product $ \mathcal{G}_{i_b} \shuffle q$ if $q$ is inserted between $i_b,i_{b{+}1}$:
\begin{equation} 
\begin{aligned}
    & F_n^{{\rm tr}(\phi^2)}(\alpha(i_1,i_2,\ldots,i_r)|i_1,\mathcal{G}_{i_1},i_2,\mathcal{G}_{i_2} \ldots,i_r, \mathcal{G}_{i_r}) \\
    =& \sum_{a,b} A(\alpha_{1},\alpha_{2},\ldots,\alpha_{a},q,\alpha_{a{+}1},\ldots,\alpha_{r} |  i_1,\mathcal{G}_{i_1},i_2 \ldots i_b,\mathcal{G}_{i_b} \shuffle q,i_{b{+}1},\ldots,i_r, \mathcal{G}_{i_r}). 
\end{aligned}
\end{equation}
For the simplest skeleton case, $\alpha(i_1,i_2,\ldots,i_r)=(i_1,i_2,\ldots,i_r)$, the result with gluons is remarkably simple:
\begin{equation}  \label{eq noreflection gluon}
\begin{aligned}
    &F_n^{{\rm tr}(\phi^2)}(i_1,i_2,\ldots,i_r| i_1,\mathcal{G}_{i_1},i_2,\mathcal{G}_{i_2},\ldots,i_r,\mathcal{G}_{i_r}) \\
    =&\sum_{1\leq a \leq b <n} A(i_1,i_2,\ldots,i_a,q,i_{a{+}1},\ldots,i_r| i_1,\mathcal{G}_{i_1},i_2,\mathcal{G}_{i_2},\ldots,i_b,\mathcal{G}_{i_b}\shuffle q,i_{b{+}1},\ldots,i_r,\mathcal{G}_{i_r}),
\end{aligned}
\end{equation}
where $q$ is shuffled with all gluons in any gluon set ${\cal G}_{i_b}$ where it needs to be inserted for the skeleton case. Let us give an example with 4 scalars explicitly:
\begin{equation}
\begin{split}
    &F_8^{{\rm tr}(\phi^2)}(1,2,5,7| 1,2,3^g,4^g,5,6^g,7,8^g) \\
    =& A(1,q,2,5,7| 1,q,2,3^g,4^g,5,6^g,7,8^g)+A(1,q,2,5,7| 1,2,(3^g,4^g) \shuffle q,5,6^g,7,8^g)  \\
    &+A(1,q,2,5,7| 1,2,3^g,4^g,5,6^g\shuffle q,7,8^g)\\
    &+A(1,2,q,5,7| 1,2,(3^g,4^g)\shuffle q,5,6^g,7,8^g)+A(1,2,q,5,7| 1,2,3^g,4^g,5,6^g\shuffle q,7,8^g) \\
    &+A(1,2,5,q,7| 1,2,3^g,4^g,5,6^g\shuffle q,7,8^g).
\end{split}
\end{equation} 
In general for $\alpha\neq \beta$ the expansion becomes more complicated but we emphasize that it is as complicated as the skeleton case; once that is given we simply shuffle $q$ with gluons in any set that it should be inserted! We simply give another example with $r=4$, which is obtained from the skeleton expansion \eqref{eq onereflection 4pt}:
\begin{equation}
\begin{split}
   &F_8^{{\rm tr}(\phi^2)}(1,2,7,5| 1,2,3^g,4^g,5,6^g,7,8^g) \\
   =& A(1,q,2,7,5 | 1,q,2,3^g,4^g,5,6^g,7,8^g) \\
   &- A(1,2,q,7,5 | 1,2,3^g,4^g,5,6^g,7,8^g \shuffle q) -A(1,2,q,7,5 | 1,2,3^g,4^g,5,6^g \shuffle q,7,8^g) \\
   &- A(1,2,7,q,5 | 1,2,(3^g,4^g) \shuffle q,5,6^g\shuffle q,7,8^g).
\end{split}
\end{equation}

We will outline a proof of this rule by checking factorizations of the expansion, but before doing so we remark that this shows how close ${\rm tr}(\phi^2)$ form factors are to amplitudes with an additional leg: once we relate all scalar diagrams contributing to the skeleton case, we simply attach gluons according to their orderings with $q$ shuffled with them, on both sides of the expansion.  
\subsubsection{Consistency checks by factorizations}

Here let us first provide a consistency check for the $\alpha(i_1,i_2,\ldots,i_r)=(i_1,i_2,\ldots,i_r)$ cases,  \eqref{eq noreflection gluon}, by analysing the factorizations on both sides, and we leave the checks for general cases in Appendix~\ref{app:proof general gluon}. Since we can choose $(i_1,\mathcal{G}_{i_1},i_2,\mathcal{G}_{i_2},\ldots,i_r,\mathcal{G}_{i_r})$ to be cyclically equivalent to $(1,2,\ldots,n)$, any possible factorization channel is governed by a planar variable $s_{c,c{+}1,\ldots,d} \rightarrow 0$. There are two possibilities for the type of particles corresponding to the cut propagator.

For a gluon propagator, it is important to notice the original form factor must factorize into a pure YM amplitude times a lower point form factor, hence $\{c,c+1,\ldots,d\}$ must be a subset of $\mathcal{G}_{i_k}$ for a certain $k\in \{1,2,\ldots,r\}$. For example, one assumes $\{c,c+1,\ldots,d\}=\{\mathcal{G}_{i_1}^\prime\} \subset \{ \mathcal{G}_{i_1} \}$ and the LHS factorizes into:
\begin{equation}\label{eq:YMSexpanfact}
    A^{\mathrm{YM}}(c,c+1,\ldots,d,I^+(\epsilon)) \frac{1}{s_{c,c+1,\ldots,d}} F_{n+c-d+2}^{{\rm tr}(\phi^2)} (i_1,i_2,\ldots,i_r| i_1,\Bar{\mathcal{G}}_{i_1}^\prime,i_2,\ldots,i_r,\mathcal{G}_{i_r}),
\end{equation}
where $\bar{\mathcal{G}}_{i_1}^\prime := (i_1+1,i_1+2,\ldots,c-1,I^-(\bar{\epsilon}),d+1,\ldots,i_2-1)$ with the intermediate gluon represented by $I^+(\epsilon),I^-(\bar{\epsilon})$. What is crucial here is that the scalar skeleton is unaffected by the factorization. In this sense, applying the expansion formula again to the lower point form factor in \eqref{eq:YMSexpanfact} gives a sum which trivially matches the factorization of the amplitudes in the RHS \eqref{eq noreflection gluon} term by term, since the amplitudes with $q$ inserted inside $\mathcal{G}_{i_1}^\prime$ do not contribute.

For a scalar propagator, the original form factor is now factorizing into a YMS amplitude times a lower point form factor.
It is easy to confirm that behaviour of both sides are equal if $\{i_1,i_r\} \not\subset  \{c,c+1,\ldots,d\}$, similar to the computation in \eqref{eq:purescalarproof}. Otherwise the analysis is more involved and we give more detailed depictions here.

For the simplicity of notations, we assume $(c,c+1,\ldots,d)=(\mathcal{G}_{i_{k}}^\prime,\ldots,i_r, \mathcal{G}_{i_{r}},i_1, \mathcal{G}_{i_{1}}^\prime)$, where $\mathcal{G}_{i_{k}}^\prime, \mathcal{G}_{i_{1}}^\prime$ are ordered subsets of $\mathcal{G}_{i_{k}},\mathcal{G}_{i_{1}}$ and are adjacent to $i_{k+1},i_1$ respectively. Hence factorizing the $n$-point form factor on the LHS of \eqref{eq noreflection gluon} gives
\begin{equation}\label{eq:noreflectiongluon2}
\begin{split}
   &A(i_{k+1},\ldots,i_r,i_1,I^+|\mathcal{G}_{i_{k}}^\prime,\ldots,i_r, \mathcal{G}_{i_{r}},i_1, \mathcal{G}_{i_{1}}^\prime,I^+) \frac{1}{s_{c,c+1,\ldots,d}} \times \\
   &\qquad F_{n+c-d+2}^{{\rm tr}(\phi^2)}(I^-,i_2,\ldots,i_k|I^-,\bar{\mathcal{G}}_{i_1}^\prime,i_2,\mathcal{G}_{i_2},\ldots,i_k,\bar{\mathcal{G}}_{i_k}^\prime),
\end{split}
\end{equation}
where and the intermediate scalar is represented by $I^+,I^-$ and $\bar{\mathcal{G}}_{i_1}^\prime :=  \mathcal{G}_{i_{1}} \backslash  \mathcal{G}_{i_{1}}^\prime $ and so does $\bar{\mathcal{G}}_{i_k}^\prime$.

Then we inspect the RHS of \eqref{eq noreflection gluon}. Among all the amplitudes being summed, we claim that the following contributions are special
\begin{equation}\label{eq:noreflectiongluon4}
    \sum_{1 \leq a < k } A(i_1,i_2,\ldots,i_a,q,i_{a{+}1},\ldots,i_k,\ldots,i_r|i_1,\mathcal{G}_{i_1},i_2,\mathcal{G}_{i_2},\ldots,i_k,\mathcal{G}_{i_k}^\prime,\Bar{\mathcal{G}}_{i_k}^\prime \shuffle q,i_{k+1},\ldots,i_r,\mathcal{G}_{i_r}),
\end{equation}
because $q$ gets to shuffle with $\Bar{\mathcal{G}}_{i_k}^\prime$. To see why this is special, we further write down its factorization on the $s_{c,c{+}1,\ldots,d}$ pole
\begin{equation}\label{eq:noreflectiongluon3}
\begin{split}
&\hskip -10pt A(i_{k+1},\ldots,i_r,i_1,I^+|\mathcal{G}_{i_{k}}^\prime,\ldots,i_r, \mathcal{G}_{i_{r}},i_1, \mathcal{G}_{i_{1}}^\prime,I^+) \frac{1}{s_{c,c+1,\ldots,d}} \times\\
    &  \Big( A(I^-,q,i_2,i_3,\ldots,i_k|I^-,\Bar{\mathcal{G}}_{i_1}^\prime,i_2,\mathcal{G}_{i_2},\ldots,i_k,\Bar{\mathcal{G}}_{i_k}^\prime \shuffle q) \  + \\
      & \ \  \sum_{2\leq a<k} A(I^-,i_2,\ldots,i_a,q,i_{a{+}1},\ldots,i_k|I^-,\Bar{\mathcal{G}}_{i_1}^\prime,i_2,\mathcal{G}_{i_2},\ldots,i_k,\Bar{\mathcal{G}}_{i_k}^\prime \shuffle q) \Big).
\end{split}
\end{equation}
Compared with \eqref{eq:noreflectiongluon2}, we observe that in that equation $q$ does not meet $\Bar{\mathcal{G}}_{i_k}^{\prime}$ if we expand the $F_{n{+}c{-}d{+}2}^{\operatorname{tr}(\phi^2)}$ therein, but \eqref{eq:noreflectiongluon3} contains $\Bar{\mathcal{G}}_{i_k}^{\prime}\shuffle q$ in the second ordering of the amplitudes. However, such a mismatch is actually not a problem because of the U(1) decoupling relation
\begin{equation*} A(I^-,q,i_2,i_3,\ldots,i_k|\gamma) +  \sum_{2\leq a<k} A(I^-,i_2,\ldots,i_a,q,i_{a{+}1},\ldots,i_k|\gamma )=0\,,
\end{equation*}
which is valid for any $\gamma$ and we pick $\gamma := (I^-,\Bar{\mathcal{G}}_{i_1}^\prime,i_2,\mathcal{G}_{i_2},\ldots,i_k,\Bar{\mathcal{G}}_{i_k}^\prime \shuffle q)$ here.


Given that the contributions from \eqref{eq:noreflectiongluon4} vanish, the rest part of the RHS of \eqref{eq noreflection gluon} is then equal to the LHS under this factorization channel, after performing the expansion for the lower point form factor. We also comment that the subtlety for $\{i_1,i_r\} \subset  \{c,c+1,\ldots,d\} $ cases considered above actually illustrate the cyclicity for \eqref{eq noreflection gluon}.


\section{A complete expansion for ${\rm tr}(F^2)$ form factors and CHY formulae} \label{sec:expand and CHY}

In this section, we show that by combining results above, we obtain an expansion of $n$-point $\mathrm{tr}(F^2)$ form factor into $(n{+}1)$-point YMS amplitudes with $r{+}1$ scalars, with coefficients given by traces of $r$ field strengths. Such an expansion reads
\begin{equation}\label{eq:general}
F_n^{{\rm tr}(F^2)}=\sum_{i_1< \cdots< i_r, r=2}^n \sum_{\alpha \in S_{r}/\mathbb{Z}_r} \mathrm{tr}^\mathrm{f}(\alpha_1,\alpha_2,\ldots,\alpha_r)\sum_{\pi \in \alpha \shuffle q } {\rm sgn}_\pi~A(\pi | 1, \cdots, q \shuffle {\cal G}_i, \cdots, n)
\end{equation}
where the first two summations come from the decomposition into ${\rm tr}(\phi^2)$ form factors with $r$ scalars, $1\leq i_1< i_2 <\cdots < i_r\leq n$, in the same ordering $\alpha$ ($\alpha_a:=\alpha(i_a)$) as the trace; the remaining part denotes the expansion of such form factors into $(n{+}1)$-point YMS amplitudes: we first sum over ordering $\pi$ for $r{+}1$ scalars with $q$ inserted into $\alpha$, and for each $\pi$ (in addition to a possible sign) we need to implicitly sum over the second ordering of all $n{+}1$ particles, with $1,2, \cdots, n$ and $q$ shuffled with gluons in some sets ${\cal G}_{i_1}, \cdots, {\cal G}_{i_r}$. The details of these two sums are given in sec. \ref{ssec:expand scalar general}. 

Since these YMS amplitudes have all been computed (see \cite{Cheung:2021zvb,Edison:2020ehu,He:2021lro} for closed-form expressions and automatized codes), our expansion thus provides an algorithm for computing $n$-point form factors explicitly. Here we summarize the number of amplitudes involved in our expression for $\mathrm{tr}(F^2)$ form factor in Table~\ref{tab:counting}. We comment that although the number of YMS amplitudes grows rather rapidly with $n$, the majority of them are bi-adjoint scalar amplitudes or those with very few gluons, while the most complicated ones ($r=2$) are relatively rare. 
\begin{table}[htbp!]
\centering
\begin{tabular}{|c|c|c|c|c|c|c|c|}
  \hline
  \diagbox{$n$}{$r$} & 2 & 3 & 4 & 5 & 6 & 7 & total \\ \hline
  3 & 4 & 3 &  &  &  &  & 7 \\ \hline
  4 & 10 & 15 & 15 &  &  &  &40  \\ \hline
  5 & 20 & 45 &  91 &  77 &    &   & 233 \\ \hline
  6 & 35 & 105 & 321 & 545  &408  &  & 1414 \\ \hline
  7 & 56 & 210 & 861 & 2198 &3293  &2210  &8828  \\
  \hline
\end{tabular}
\caption{The number of $r+1$ scalar YMS amplitudes appear in the decomposition of $F_n^{{\rm tr}(F^2)}(1,2,\ldots,n)$.}
\label{tab:counting}
\end{table}

\subsection{CHY formulae for form factors}

Another interesting consequence of \eqref{eq:general} is a CHY formula for 
the ${\rm tr}(F^2)$ form factor itself (as well as all ${\rm tr}(\phi^2)$ form factors as intermediate steps). This simply arise from CHY formulas for YMS amplitudes, which we recall that for $r{+}1$ scalars in ordering $\pi$ and all $n{+}1$ particles in ordering $\beta(1,\cdots, n{+}1)$ read
\begin{equation}
A_{n{+}1}(\pi|\beta)=\int d\mu_{n{+}1}~{\rm PT}_{n{+}1} (\beta)~{\rm PT}_{r{+}1} (\pi) {\rm Pf}\Psi_{n{-}r}
\end{equation}
where the ingredients, such as the measure, the Parke-Taylor factors (of length $n{+}1$ and $r{+}1$), and the Pfaffian (of remaining $n{-}r$ gluons) are reviewed in Appendix~\ref{app:CHY} for completeness. 

The integrand for this general CHY formula takes the nice form
\begin{equation}
{\rm PT}(1,\cdots, n)\sum_{i_1< \cdots< i_r, r=2}^n {\rm Pf} \Psi_{n{-}r}~\sum_{\alpha \in S_{r}/\mathbb{Z}_r} \mathrm{tr}^\mathrm{f}(\alpha_1,\alpha_2,\ldots,\alpha_r)\sum_{\pi \in \alpha \shuffle q } {\rm sgn}_\pi~{\rm PT}(\pi) \times {\cal S}_{\pi}
\end{equation}
where we have factored out an overall ${\rm PT}(1,\cdots,n)$ and defined the ``inverse soft factor" ${\cal S}_{\pi}$ for inserting $q$ into gluon sets as indicated by the ordering $\pi$,
\begin{equation}\label{ISF}
{\cal S}_{\pi}:=\sum_{a \in I(\pi)} \Big(\frac 1 {\sigma_{i_a, q}}- \frac 1 {\sigma_{i_{a{+}1}, q}}\Big)\,,
\end{equation}
where $\sigma_{i_a,q} := \sigma_{i_a} - \sigma_q$. Note that when $q$ is shuffled with a gluon set ${\cal G}_{i_1}$ between scalars $i_1$ and $i_2$, the overall effect is nothing but the factor involving the two end points: $\sum_{j=i_1}^{i_2{-}1} \frac 1 {\sigma_{j,q}}- \frac 1{\sigma_{j{+}1, q}}=\frac 1 {\sigma_{i_1, q}}- \frac 1 {\sigma_{i_2, q}}$; ${\cal S}_\pi$ is a sum of such factors for those gluon sets with $q$ inserted (as indicated by $I(\pi) \subset \{1, \cdots, r\}$). Note that if we sum over all gluon sets $I(\pi)=\{1, \cdots, r\}$, it vanishes because of the U(1) identity. 

For example, for $r=2$, $\alpha=(i, j)$ with $i<j$, there is only one $\pi=i, q, j$, and the the CHY integrand is nothing but 

\begin{equation}
    \mathrm{PT}(1,2,\ldots,n) \operatorname{Pf} \Psi_{n{-}2} (\overline{i,j}) \ \mathrm{tr}^\mathrm{f}(i,j) \mathrm{PT}(i,q,j)  \Big(\frac{1}{\sigma_{i,q}}-\frac{1}{\sigma_{j,q}}\Big),
\end{equation}
For $r=3$ with $\alpha=(i,j,k)$ where $i<j<k$, we sum over two possible orderings $\pi=(i, q, j,k)$ and $(i, j, q, k)$, and the corresponding soft factors are simply 
\begin{equation}
    {\cal S}_{i,q,j,k}=\Big(\frac{1}{\sigma_{i,q}}-\frac{1}{\sigma_{j,q}}\Big)+ \Big(\frac{1}{\sigma_{j,q}}-\frac{1}{\sigma_{k,q}}\Big)=\Big(\frac{1}{\sigma_{i,q}}-\frac{1}{\sigma_{k,q}}\Big), \quad {\cal S}_{i,j,q,k}= \Big(\frac{1}{\sigma_{j,q}}-\frac{1}{\sigma_{k,q}}\Big). 
\end{equation}
Let us finally give some more examples for $r=4$. For  $\alpha=(i,j,k,l)$ with $i<j<k<l$, the factors are
\begin{equation}
    {\cal S}_{i,q,j,k,l}= \Big(\frac{1}{\sigma_{i,q}}-\frac{1}{\sigma_{l,q}}\Big), \quad {\cal S}_{i,j,q,k,l}= \Big(\frac{1}{\sigma_{j,q}}-\frac{1}{\sigma_{l,q}}\Big), \quad {\cal S}_{i,j,k,q,l}= \Big(\frac{1}{\sigma_{k,q}}-\frac{1}{\sigma_{l,q}}\Big)\,.
\end{equation}
For $\alpha=(i,j,l,k)$ the soft factors become
\begin{align}
    &{\cal S}_{i,q,j,l,k}= \Big(\frac{1}{\sigma_{i,q}}-\frac{1}{\sigma_{j,q}}\Big) \quad  {\cal S}_{i,j,q,l,k}=\Big(\frac{1}{\sigma_{k,q}}-\frac{1}{\sigma_{l,q}}\Big) +\Big(\frac{1}{\sigma_{l,q}}-\frac{1}{\sigma_{i,q}}\Big)=\Big(\frac{1}{\sigma_{k,q}}-\frac{1}{\sigma_{i,q}}\Big) \nonumber \\
    & {\cal S}_{i,j,l,q,k}= \Big(\frac{1}{\sigma_{k,q}}-\frac{1}{\sigma_{l,q}}\Big).
\end{align}
Note that for this case, the ${\rm sgn}_\pi$'s are ${\rm sgn}_{i,q,j,l,k}=1$ and ${\rm sgn}_{i,j,q,l,k}={\rm sgn}_{i,j,l,q,k}=-1$. In fact, for $\alpha=(i_1,i_2,\ldots,i_r)$ with $i_1<i_2<\ldots<i_r$ which corresponds to the $\alpha=\beta$ case of the scalar skeleton, the result is remarkably simple: there are $(r{-}1)$ possible orderings $\pi=(i_1,i_2,\ldots,i_a,q,i_{a{+}1},\ldots,i_r)$ for $a=1,2,\ldots,r{-}1$, each accompanied with ${\rm sgn}_\pi=+1$, and the inverse soft factors are
\begin{equation}\label{specialISF}
    {\cal S}_{i_1,i_2,\ldots,i_a,q,i_{a{+}1},\ldots,i_r}=\frac{1}{\sigma_{i_a,q}}-\frac{1}{\sigma_{i_r,q}} .
\end{equation}

We remark that, as for any $(n{+}1)$-point amplitude in this paper, our CHY formula needs to be evaluated with the prescription $q\to - \sum_{i=1}^n p_i$. This can be directly realized if we use the partial SL$(2,\mathbb{C})$ gauge fixing with $\sigma_q \to \infty$, which eliminate $q$ from the formula.


We have recorded the expansion of ${\rm tr}(F^2)$ form factors into YMS amplitudes up to $n=9$ in the auxiliary {\sc Mathematica} file. We have also provided the explicit result up to $n=7$ where the YMS amplitudes are evaluated by the package in~\cite{He:2021lro} based on CHY formula. As a simple illustration, we give explicit results for $\mathrm{tr}(F^2)$ form factor for $n=3,4$. For example, combining \eqref{eq:3pttrphi2} and \eqref{eq:3pttrF2b}, one gets
\begin{equation}
    F_3^{{\rm tr}(F^2)}(1,2,3)=\frac{1}{s_{1,2}}\left(-p_1\cdot \epsilon_2 \mathrm{tr}^\mathrm{f}(1,3) +\mathrm{tr}^\mathrm{f}(2,3) p_2\cdot \epsilon _1+2 \mathrm{tr}^\mathrm{f}(1,2,3)\right)+\mathrm{cyclic}(1,2,3).
\end{equation}
The $n=4$ result involves $10$ graphs organized into cyclic orbits of length $4$ and $2$:
\begin{equation}
\begin{aligned}
    &F_4^{{\rm tr}(F^2)}(1,2,3,4) \\
    =& \frac{1}{s_{2,3} s_{2,3,4}}( \mathrm{tr}^\mathrm{f}(1,2)( p_2\cdot \epsilon _3 p_2\cdot \epsilon _4+ p_2\cdot \epsilon _3 p_3\cdot \epsilon _4)-\mathrm{tr}^\mathrm{f}(1,3)( p_2\cdot \epsilon _4 p_3\cdot \epsilon _2+ p_3\cdot \epsilon _2 p_3\cdot \epsilon _4)\\
    & +\mathrm{tr}^\mathrm{f}(1,4)( -p_2\cdot \epsilon _3 p_4\cdot \epsilon _2+ \epsilon _2\cdot f_3\cdot p_4)-2 \mathrm{tr}^\mathrm{f}(1,2,3)( p_2\cdot \epsilon _4+ p_3\cdot \epsilon _4)-2 \mathrm{tr}^\mathrm{f}(1,2,4) p_2\cdot \epsilon _3\\
    &+2 \mathrm{tr}^\mathrm{f}(1,3,4) p_3\cdot \epsilon _2+2 \mathrm{tr}^\mathrm{f}(1,2,3,4)-2 \mathrm{tr}^\mathrm{f}(1,3,2,4))\\
    &+ \frac{1}{s_{3,4} s_{2,3,4}}(\mathrm{tr}^\mathrm{f}(1,2) (p_2\cdot \epsilon _3 p_3\cdot \epsilon _4- \epsilon _3\cdot f_4\cdot p_2)-\mathrm{tr}^\mathrm{f}(1,3) (p_3\cdot \epsilon _2 p_3\cdot \epsilon _4+ \epsilon _2\cdot f_4\cdot p_3)\\
    &+\mathrm{tr}^\mathrm{f}(1,4) (p_4\cdot \epsilon _2 p_4\cdot \epsilon _3+ \epsilon _2\cdot f_3\cdot p_4)-2 \mathrm{tr}^\mathrm{f}(1,2,3) p_3\cdot \epsilon _4+2 \mathrm{tr}^\mathrm{f}(1,2,4) p_4\cdot \epsilon _3\\
    &+2 \mathrm{tr}^\mathrm{f}(1,3,4) (p_3\cdot \epsilon _2+ p_4\cdot \epsilon _2)+2 \mathrm{tr}^\mathrm{f}(1,2,3,4)-2 \mathrm{tr}^\mathrm{f}(1,2,4,3))\\
    &+\frac{1}{2 s_{1,2} s_{3,4}}(\mathrm{tr}^\mathrm{f}(1,3) p_1\cdot \epsilon _2 p_3\cdot \epsilon _4-\mathrm{tr}^\mathrm{f}(1,4) p_1\cdot \epsilon _2 p_4\cdot \epsilon _3-\mathrm{tr}^\mathrm{f}(2,3) p_2\cdot \epsilon _1 p_3\cdot \epsilon _4\\
    &+\mathrm{tr}^\mathrm{f}(2,4) p_2\cdot \epsilon _1 p_4\cdot \epsilon _3-2 \mathrm{tr}^\mathrm{f}(1,2,3) p_3\cdot \epsilon _4+2 \mathrm{tr}^\mathrm{f}(1,2,4) p_4\cdot \epsilon _3-2 \mathrm{tr}^\mathrm{f}(1,3,4) p_1\cdot \epsilon _2\\
    &+2 \mathrm{tr}^\mathrm{f}(2,3,4) p_2\cdot \epsilon _1+2 \mathrm{tr}^\mathrm{f}(1,2,3,4)-2 \mathrm{tr}^\mathrm{f}(1,2,4,3)) + \mathrm{cyclic}(1,2,3,4).
\end{aligned}
\end{equation}

\section{Conclusion and Outlook}

In this paper we have presented two types of new relations for tree-level form factors and scattering amplitudes in Yang-Mills-scalar theory. Not only do we have a decomposition of ${\rm tr}(F^2)$ form factors into ${\rm tr}(\phi^2)$ ones similar to the so-called universal expansion for amplitudes~\cite{Dong:2021qai}, but these $n$-point form factors can also be further expanded into $(n{+}1)$-point YMS amplitudes with an additional scalar leg and unity coefficients. As we have seen, such new relations provide an efficient method for computing form factors in terms of YMS amplitudes, which had been computed mainly using Feynman diagrams. Moreover, combined with CHY formulae and even closed-form expressions for YMS amplitudes, we have obtained such formulae for all-multiplicity form factors in general dimension. 

There are numerous open questions raised by our preliminary investigations. We have only considered form factors with length-two operators and it would be highly desirable to see if such relations, especially the first type, exist for form factors with other operators, such as ${\rm tr}(F^3)$. It would also be interesting to understand in general why such universal expansions exist, perhaps from certain ``uniqueness" theorem for form factors similar to amplitudes \cite{Arkani-Hamed:2016rak,Rodina:2016jyz,Rodina:2016mbk}. Moreover, already the simplest expansion into amplitudes for ${\rm tr}(\phi^2)$ form factor with two adjacent scalars, \eqref{eq:prototype}, provide an explanation for ``double-copy" relations found in~\cite{Lin:2021pne}, and an important question is whether these new relations, which connect form factors to amplitudes, will be useful to study double copy of $F^{{\rm tr}(\phi^2)}$ with more than two scalars and even $F^{{\rm tr}(F^2)}$. 

The existence of CHY formulae for form factors is also very suggestive, and we expect to find more simplifications, along the line of \eqref{ISF} and \eqref{specialISF}, as well hidden structures in such formulae. It would be interesting to see if they can be derived from certain correlators in ambitwistor or conventional string theories~\cite{Mason:2013sva, Berkovits:2013xba}. Besides, it would be interesting to understand the aforementioned form-factor double copy from the viewpoint of the CHY formalism, which should be particularly suitable for studying such double-copy relations.  

Last but not least, given the success at tree level, it is natural to ask if similar relations can be found for form factors at loop level, {\it e.g.} if loop integrands of $F^{{\rm tr}(F^2)}$ can be expanded into those of $F^{{\rm tr}(\phi^2)}$, and even related to integrands for amplitudes. Such explorations may allow us to extend even more fascinating structures found for multi-loop amplitudes into form factors.

\begin{acknowledgments}
We would like to thank Gang Chen, Congkao Wen, Yong Zhang, Mao Zeng and especially Gang Yang for helpful discussions. The research of S. H. is supported in part by the Key Research Program of CAS, Grant No. XDPB15 and National Natural Science Foundation of China, Grants No. 11935013, No. 11947301, No. 12047502, and No. 12047503. G.L. is supported in part by the National Natural Science Foundation of
China Grants No.~11935013. G.L. thanks the Higgs Centre for Theoretical Physics at the University of Edinburgh for the Visiting Researcher Scheme. G.L. also thanks the Queen Mary University for hospitality in the final stage of the paper. 

\end{acknowledgments}

\newpage

\appendix
\section{Details in the proof of $\operatorname{tr}(F^2)$ form factor decomposition} \label{app:proof dec trF2}

In this appendix, we complete the details in the proof of the $\operatorname{tr}(F^2)$ decomposition, especially regarding the factorization in \eqref{eq:cutderivative1}. 

From unitarity, the residue in \eqref{eq:cutderivative1} is nothing but a lower point Yang-Mills form factor and a Yang-Mills amplitudes
\begin{equation}\label{eq:trF2factorization}
    \text{Res}\big[F_n^{\operatorname{tr}(F^2)}(1,\ldots,n)\big]_{s_{1,\ldots,m}= 0}=A^{ \rm YM}(1,\ldots,m,I^{+}(\epsilon)){\scriptstyle \circ} F_{n{-}m{+}1}^{\operatorname{tr}(F^2)}(I^{-}(\bar{\epsilon}),m{+}1,\ldots,n)\,.
\end{equation}
Based on this factorization form, we explain why \eqref{eq:splitderivative}, that is ``closed-trace" factorizing into ``open-trace$\times$closed-trace", is correct. We re-write \eqref{eq:splitderivative} here
\begin{equation}\label{eq:splitderivativeb}
    \mathcal{D}_{(i_1,\ldots,i_r)}\cong \widetilde{\mathcal{D}}_{(i_1,\ldots,i_{r^{\prime}},I^{+}(\epsilon))}{\scriptstyle \circ} {\mathcal{D}}_{(I^{-}(\bar{\epsilon}),i_{r^{\prime}+1},\ldots,i_r)}\,,
\end{equation}
where the ${\scriptstyle \circ}$ implies a summation over helicity and the $\cong$ symbol means that the two sides of \eqref{eq:splitderivativeb} are equal on support of the factorization \eqref{eq:trF2factorization}.

On the RHS of \eqref{eq:trF2factorization}, we encounter the following form in general
\begin{equation}
    \epsilon_{\mu}(I^{+})\bigg(\sum_{\alpha}f_{\alpha}(\{p_{i}\})M^{\mu}_{\alpha}(\{\epsilon_{i}\};\{p_{i}\})\bigg)\bar{\epsilon}_{\nu}(I^{-})\bigg(\sum_{\beta}f_{\beta}(\{p_{j}\})M^{\nu}_{\beta}(\{\epsilon_{j}\};\{p_{j}\})\bigg)\,,
\end{equation}
where $i\in\{1,\ldots,m\}$, $j\in\{m{+}1,\ldots,n\}$, $f$ are rational functions of Mandelstum variables and $M$ are monomials as functions linearly depending on all the polarizations in its argument. It is straight forward to check the following relation holds
\begin{equation}\label{eq:trF2temp1}
    f_{\alpha}(\{p_i\})=  \epsilon_{\mu}(I^{+})\bigg(\sum_{\alpha}f_{\alpha}(\{p_{i}\})M^{\mu}_{\alpha}(\{\epsilon_{i}\};\{p_{i}\})\bigg)\Big|_{\epsilon_{\mu}(I^{+})M^{\mu}_{\alpha}(\{\epsilon_{i}\};\{p_{i}\})}\,,
\end{equation}
where $|_{\rm T}$ means extracting the $\rm T$ coefficients. 
This process is equivalent to taking derivatives with respect to $\epsilon_{\mu}(I^{+})M^{\mu}_{\alpha}(\{\epsilon_{i}\};\{p_{i}\})$. Similarly, we have a derivative to get $f_{\beta}(\{p_j\})$, and we can get $f_{\alpha}(\{p_j\})f_{\beta}(\{p_j\})$ with
\begin{align}\label{eq:trF2temp2}
     f_{\alpha}(\{p_i\}) f_{\beta}(\{p_j\})= & \epsilon_{\mu}(I^{+})\bigg(\sum_{\alpha}f_{\alpha}(\{p_{i}\})M^{\mu}_{\alpha}(\{\epsilon_{i}\};\{p_{i}\})\bigg)\Big|_{\epsilon_{\mu}(I^{+})M^{\mu}_{\alpha}(\{\epsilon_{i}\};\{p_{i}\})}\\
     & \times  \bar{\epsilon}_{\nu}(I^{-})\bigg(\sum_{\beta}f_{\beta}(\{p_{j}\})M^{\nu}_{\beta}(\{\epsilon_{j}\};\{p_{j}\})\bigg)\Big|_{\bar{\epsilon}_{\nu}(I^{-})M^{\nu}_{\beta}(\{\epsilon_{j}\};\{p_{j}\})}\nonumber \,.
\end{align}
Then the important thing to notice is that after the polarization sum, we can also extract the $f_{\alpha}(\{p_j\})f_{\beta}(\{p_j\})$ piece by
\begin{align}\label{eq:trF2temp3}
    f_{\alpha}(\{p_{i}\})&f_{\beta}(\{p_j\})=\\
    \nonumber&\bigg(\sum_{\alpha,\beta}f_{\alpha}(\{p_{i}\})M^{\mu}_{\alpha}(\{\epsilon_{i}\};\{p_{i}\})f_{\beta}(\{p_{j}\})M^{\mu}_{\beta}(\{\epsilon_{j}\};\{p_{j}\})\bigg)\Big|_{M^{\mu}_{\alpha}(\{\epsilon_{i}\};\{p_{i}\})M_{\mu,\beta}(\{\epsilon_{j}\};\{p_{j}\})}\,.
\end{align}
Once again, $|$ represents extracting coefficients equivalent to taking derivatives. 
Comparing \eqref{eq:trF2temp2} and  \eqref{eq:trF2temp3} we have 
\begin{equation}\label{eq:splitderivative2}
    D[M^{\mu}_{\alpha}(\{\epsilon_{i}\};\{p_{i}\})M_{\mu,\beta}(\{\epsilon_{j}\};\{p_{j}\})]\cong D[\epsilon_{\mu}(I^{+})M^{\mu}_{\alpha}(\{\epsilon_{i}\};\{p_{i}\})]{\scriptstyle \circ} D[\bar{\epsilon}_{\nu}(I^{-})M^{\nu}_{\alpha}(\{\epsilon_{j}\};\{p_{j}\})]\,.
\end{equation}
where $D$ is a general notation for derivatives and we mention again that $\cong$ means equal on the support of factorization. 
As a result, an expression like \eqref{eq:splitderivativeb} seems to be true.

To prove \eqref{eq:splitderivativeb}, we just need to be particularly careful about the ``boundaries" highlighted in red. From \eqref{eq:splitderivative2}, we get 
\begin{equation}
    \mathcal{D}_{(i_1,\ldots,i_r)}=\partial_{({\color{red}\bm \epsilon_{i_1}}\cdot p_{i_2})}\cdots \partial_{(\epsilon_{i_r}\cdot {\color{red}\bm p_{i_1}})}\cong \partial_{(\epsilon_{i_1}\cdot p_{i_2})}\cdots \partial_{(\epsilon_{i_{r^{\prime}}}\cdot \epsilon_{I^{+}})}{\scriptstyle \circ} \partial_{(\bar{\epsilon}_{I^{-}}\cdot p_{i_{r^{\prime}+1}})}\cdots \partial_{(\epsilon_{i_{r}}\cdot {\color{red}\bm p_{i_1}})}\,.
\end{equation}
To reproduce \eqref{eq:splitderivativeb}, however, the last term should be $\partial_{(\bar{\epsilon}_{I^{-}}\cdot p_{i_{r^{\prime}+1}})}\cdots \partial_{(\epsilon_{i_{r}}\cdot {\color{red}\bm p_{I^{-}}})}$
rather than $\partial_{(\bar{\epsilon}_{I^{-}}\cdot p_{i_{r^{\prime}+1}})}\cdots \partial_{(\epsilon_{i_{r}}\cdot {\color{red}\bm p_{i_1}})}$. 
Fortunately, these two derivatives are equivalent when acting on the second block $F^{\operatorname{tr}(F^2)}$ in \eqref{eq:trF2factorization}. The reason is that by momentum conservation $p_{I^{-}}=\sum_{i=1}^{m}p_i$ and $p_{I^{-}}$ is the only way that $p_1$ can appear in the second block, so 
\begin{equation}
    \partial_{(\epsilon_{i_{t}}\cdot {\color{red}\bm p_{i_1}})}F_{n{-}m{+}1}^{\operatorname{tr}(F^2)}({\color{red} I^{-}(\bar{\epsilon})},m{+}1,\ldots,n)=\partial_{(\epsilon_{i_{t}}\cdot {\color{red}\bm p_{I^{-}}})}F_{n{-}m{+}1}^{\operatorname{tr}(F^2)}({\color{red} I^{-}(\bar{\epsilon})},m{+}1,\ldots,n)\,,
\end{equation}
and \eqref{eq:splitderivativeb} is proved.

\section{Consistency of the  $\mathrm{tr}(\phi^2)$ form factor expansion}

We show the process introduced in sec.~\ref{sec:phi2decomp} of expanding $F_n^{{\rm tr}(\phi^2)}(\alpha(i_1,i_2,\ldots,i_r)|i_1,\mathcal{G}_{i_1},i_2,$ $\mathcal{G}_{i_2}, \ldots,i_r, \mathcal{G}_{i_r})$ on YMS amplitudes is indeed consistent. 
In the main text, we have already addressed the case with the simplest scalar skeleton that is $\alpha(i_1,i_2,\ldots,i_r)=(i_1,i_2,\ldots,i_r)$. In this appendix we focus on some details with more general scalar skeleton. 
As in sec.~\ref{sec:phi2decomp}, we will first discuss the pure scalar cases $F_{r}^{\operatorname{tr}(\phi^2)}(\alpha|\beta)$ whose expansion is given in sec.~\ref{ssec:scalarskeleton}. In \ref{app:equivalencea} and \ref{app:equivalence}, we clarify that the expansion process of the $\alpha\neq\beta$ case is free of some choices which seem to be artificial.
On support of the equivalence among those choices, in \ref{app:unitarity} we show that the expansion agrees with the requirement of unitarity. Furthermore, in \ref{app:proof general gluon} we argue that our rule of inserting gluons into general scalar skeletons in sec.~\ref{ssec:withgluons} produces result with correct factorizations behaviour.


\subsection{Equivalence (I)}\label{app:equivalencea}
We begin with two basic observations regarding the cyclicity and reflection property of our algorithm for the $\alpha\neq \beta$ case, parallel to the argument for the $\alpha= \beta$ case.

Based our reduction algorithm, one only needs to worry about the cyclicity in the Step (4) of the algorithm. 
Before going on, we introduce an ``arrow" notation: 
we represent particle $q$ in the first ordered set by a down black arrow, while two down blue arrows in the second ordered set imply a summation with $q$ allowed to be inserted between these arrows(including the boundaries), {\it e.g.}
\begin{equation*}
\begin{aligned}
    & A((8,1,2),\downarrow,(6,7),(3,4,5) | (3,4,5),{\color{blue}\bm{\downarrow,(6,7),\downarrow}} ,(8,1,2) ) \\
    & :=A( (8,1,2),q,(6,7),(3,4,5) | (3,4,5),(6,7) \shuffle q ,(8,1,2) )
\end{aligned}
\end{equation*}
and we also have 
\begin{equation*}
\begin{aligned}
    & A((8,1,2),\downarrow,(6,7),(3,4,5) | (3,4,5),{\color{blue}\bm{(6,7),\downarrow}} ,(8,1,2) ) \\
    & :=A( (8,1,2),q,(6,7),(3,4,5) | (3,4,5),(6,q,7),(8,1,2) )\\
    &\ \ \  +A( (8,1,2),q,(6,7),(3,4,5) | (3,4,5),(6,7),q,(8,1,2) )
\end{aligned}
\end{equation*}
if there is no boundary term $A( (8,1,2),q,(6,7),(3,4,5) | (3,4,5),q,(6,7),(8,1,2) ) $ on the RHS. 

We denote the contribution from the Step (4) in sec.~\ref{ssec:expand scalar general} by $\widetilde{F}_t^{\mathrm{tr}(\phi^2)}(I_1 ,\ldots,I_t |I_1 ^{\prime},\ldots,I_t ^{\prime})$, which is an effective $t$-point form factor. It is important to note that this effective form factor is proportional to $\prod_{i=1}^{t} s_{I_i}^{-1}$, which can be easily proved by considering the residue of $s_{\mathcal{I}_{i}q}$ where $\mathcal{I}_{i}$ is non-empty proper subset
of $I_i$ for $i=1,2,\ldots,t$. On the support of this fact, we prove that its expansion is equivalent to that of $\widetilde{F}_t^{\mathrm{tr}(\phi^2)}(I_2 ,\ldots,I_1 |I_2 ^{\prime},\ldots,I_1 ^{\prime})$.
The difference between these two contributions is
\begin{align}\label{eq:cyclicitydiff}
    &-A(I_1 ,\downarrow,I_2 ,\ldots,I_t |I_1 ^{\prime},{\color{blue}\bm{\downarrow,I_2 ^{\prime},\ldots,\downarrow}},I_t ^{\prime})\\
    &+\sum_{a=2}^{t{-}1}A(I_2 ,\ldots,I_{a} ,\downarrow,I_{a{+}1} ,I_t ,I_1 |I_2 ^{\prime},\ldots,{\color{blue}\bm{I_t ^{\prime},\downarrow}},I_1 ^{\prime})+A(I_1 ,I_2 ,\ldots,I_t ,\downarrow|I_1 ^{\prime},I_2 ^{\prime},\ldots,I_t ^{\prime},\downarrow)\,, \nonumber
\end{align}
which can be shown to vanish by analysing the involved Feynman diagrams as follows. 

From the discussions in sec.~\ref{sec:phi2decomp}, we know that $I_1,\ldots,I_t$ are subgroups so that every Feynman diagram in $\widetilde{F}_t^{\mathrm{tr}(\phi^2)}(I_1 ,\ldots,I_t |I_1 ^{\prime},\ldots,I_t ^{\prime})$, containing the $s_{I_1}\cdots s_{I_t}$ propagators, looks like
\begin{equation}\label{eq:Feynmandiff}
   \begin{aligned}
       \includegraphics[width=0.22\linewidth]{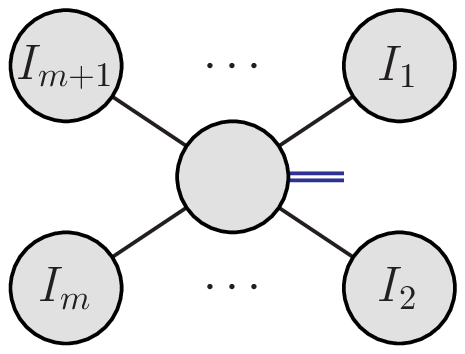}
   \end{aligned}\,.
\end{equation}
Now we would like to argue that the contributions in the first line in \eqref{eq:cyclicitydiff} involve only a special type of Feynman diagram after performing the summation
\begin{equation}\label{eq:Feynmandiffrem}
    A(I_1 ,\downarrow,I_2 ,\ldots,I_t |I_1 ^{\prime},{\color{blue}\bm{\downarrow,I_2 ^{\prime},\ldots,\downarrow}},I_t ^{\prime})=\begin{aligned}
       \includegraphics[width=0.22\linewidth]{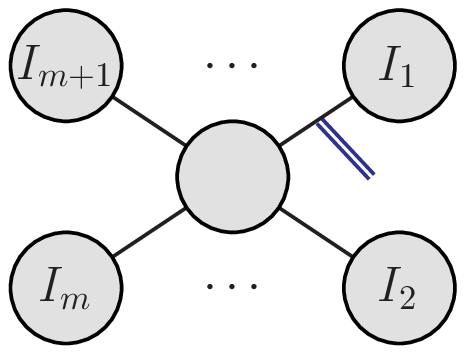}
   \end{aligned}
    \,,
\end{equation}
For any given Feynman diagram looking like \eqref{eq:Feynmandiff}, we will show that they can not appear in the sum $ A(I_1 ,\downarrow,I_2 ,\ldots,I_t |I_1 ^{\prime},{\color{blue}\bm{\downarrow,I_2 ^{\prime},\ldots,\downarrow}},I_t ^{\prime})$, unless it takes the form of the RHS of \eqref{eq:Feynmandiffrem}. This is because the same diagram appear in two double ordered amplitudes with an opposite sign as
\begin{align}
   &A(I_1 ,\downarrow,I_2 ,\ldots,I_t |I_1 ^{\prime},{{\downarrow,I_2 ^{\prime},\ldots}},I_t ^{\prime})= \cdots + \hskip -5pt \begin{aligned}
       \includegraphics[width=0.22\linewidth]{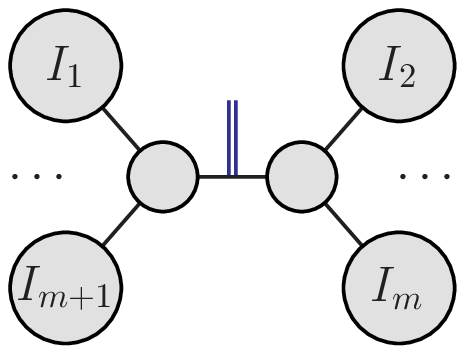}
   \end{aligned} + \cdots\\
   &A(I_1 ,\downarrow,I_2 ,\ldots,I_t |I_1 ^{\prime},\ldots,I_m ^{\prime},\downarrow,I_{m{+}1} ^{\prime},\ldots,I_t ^{\prime})= \cdots - \hskip -5pt \begin{aligned}
       \includegraphics[width=0.22\linewidth]{fig/subgroupFeynmans3.eps}
   \end{aligned} + \cdots \nonumber
\end{align} 
As a result, when considering \eqref{eq:Feynmandiffrem}, the summation over all the double ordered amplitudes gives only the special type of diagram therein which can not be cancelled. 
Direct calculations of many examples also corroborates \eqref{eq:Feynmandiffrem}. Similarly, the second row in \eqref{eq:cyclicitydiff} gives the same diagrams so that \eqref{eq:cyclicitydiff} is actually zero.

Another point we would like to make is the reflection property in Step (2) and (3). Using the same example in the Step (3), the reflection property means  that the following identity holds
{\small
\begin{equation}
\begin{aligned}
    &A(\ldots,{\overline{I_1 }},\downarrow,I_2 ,\underline{I}_3 ,\ldots|\ldots,\overline{I}_1 ,{\color{blue}\bm{\downarrow,I_2^{\prime} ,\downarrow}},\underline{I_3 },\ldots)+A(\ldots,\overline{I_1 },I_2 ,\downarrow,\underline{I_3 },\ldots|\ldots,\overline{I_1^{\prime}},I_2^{\prime} ,\downarrow,\underline{I_3^{\prime} },\ldots)\\
    =&A(\ldots,\underline{I_1 },I_2 ,\downarrow,\overline{I_3 },\ldots|\ldots,\underline{I_1^{\prime} },{\color{blue}\bm{\downarrow,I_2^{\prime} ,\downarrow}},\overline{I_3^{\prime} },\ldots)+A(\ldots,\underline{I_1 },\downarrow,I_2 ,\overline{I_3 },\ldots|\ldots,\underline{I_1^{\prime} },\downarrow,I_2^{\prime} ,\overline{I_3^{\prime} },\ldots)\,,
\end{aligned}
\end{equation}}
where we intentionally use the overline/underline to specify the first/last elements in the sub-ordering. 
It is suffice to show that
{\small
\begin{equation}\label{eq:phi2temp1}
    A(\ldots,I_1 ,\downarrow,I_2 ,I_3 ,\ldots|\ldots,I_1 ,{\color{blue}\bm{I_2 ,\downarrow}},I_3 ,\ldots)=A(\ldots,I_1 ,I_2 ,\downarrow,I_3 ,\ldots|\ldots,I_1 ,{\color{blue}\bm{\downarrow,I_2 }},I_3 ,\ldots)\,.
\end{equation}}
Similar to the above discussion on cyclicity, we know that on both sides we have poles $s_{I_2}s_{q I_2}$ so that the Feynman diagrams on both sides can only be $\begin{aligned}
    \includegraphics[width=0.17\linewidth]{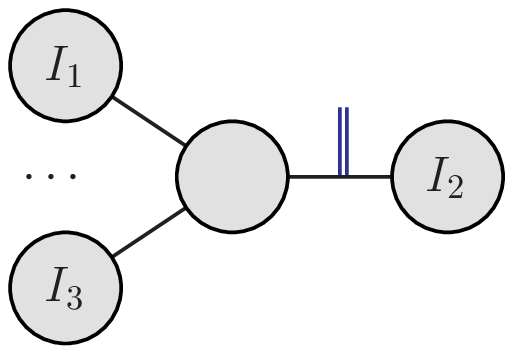}
\end{aligned}\,,$
which is why \eqref{eq:phi2temp1} is valid. 

\subsection{Equivalence (II)} \label{app:equivalence}
In the 8-point example in sec.~\ref{ssec:scalarskeleton}, we already give two different approaches of reduction shown in Figure.~\ref{fig:8pt example1} and Figure.~\ref{fig:8pt example2}. From the explicit results, we see that those two are equivalent, and we will prove a similar conclusion in general in this section. 
When considering the mutual partial triangulation diagram, our algorithm in sec.~\ref{ssec:scalarskeleton} is basically cutting out blocks separated by diagonals from the polygon successively. Now we show that different orderings of these cutting out operations give the same result in the end.

Let us take any diagonal line in the polygon and define the two blocks on either sides of the line as $P_{L}$ and $P_{R}$.
There are several possibilities for the ``final" contribution, which is the Step (4) of our algorithm, as illustrated in Figure.\ref{fig:consistency1}, where the last polygon with no diagonal is painted orange:
\begin{figure}[htbp!]
	\centering
	\subfloat[]{
		\begin{minipage}[c]{0.24\linewidth}
			\centering
			\begin{tikzpicture}
				\draw (140:1.5cm) arc(140:450:1.5cm); 	\draw[orange] (90:1.5cm) arc(90:140:1.5cm) ;
				\draw (90:1.5cm) -- (270:1.5cm); 
				\draw[orange] (90:1.5cm) -- (140:1.5cm);  \draw (140:1.5cm) -- (190:1.5cm); 
				\node[right] at (205:1.5cm) {$\ldots$};
				\node[above] at (115:1.5cm) {\small $I_1$}; \node[left] at (165:1.4cm) {\small $I_2$};
				\node[below] at (245:1.5cm) {\small $I_s$};
				\draw (220:1.5cm) -- (270:1.5cm); 
	            \draw (90:1.5cm) -- (40:1.5cm); 
				\draw (40:1.5cm) -- (350:1.5cm);  \draw (320:1.5cm) -- (270:1.5cm); 
				\node[left] at (335:1.5cm) {$\ldots$};
				\node[above] at (65:1.5cm) {\small $J_1$}; \node[right] at (15:1.4cm) {\small $J_2$};
				\node[below] at (295:1.5cm) {\small $J_t$};
				\node[left=0.25cm] at (0,0) {$P_L$};  	\node[right=0.25cm] at (0,0) {$P_R$}; 
			\end{tikzpicture}
		\end{minipage}
	}
	\subfloat[]{
		\begin{minipage}[c]{0.24\linewidth}
			\centering
			\begin{tikzpicture}
				\draw (350:1.5cm) arc(350:40:1.5cm); 	\draw[orange] (40:1.5cm) arc(40:-10:1.5cm) ;
				\draw (90:1.5cm) -- (270:1.5cm); 
				\draw (90:1.5cm) -- (140:1.5cm);  \draw (140:1.5cm) -- (190:1.5cm); 
				\node[right] at (205:1.5cm) {$\ldots$};
				\node[above] at (115:1.5cm) {\small$I_1$}; \node[left] at (165:1.4cm) {\small$I_2$};
				\node[below] at (245:1.5cm) {\small$I_s$};
				\draw (220:1.5cm) -- (270:1.5cm); 
	            \draw (90:1.5cm) -- (40:1.5cm); 
				\draw[orange] (40:1.5cm) -- (350:1.5cm);  \draw (320:1.5cm) -- (270:1.5cm); 
				\node[left] at (335:1.5cm) {$\ldots$};
				\node[above] at (65:1.5cm) {\small$J_1$}; \node[right] at (15:1.4cm) {\small$J_2$};
				\node[below] at (295:1.5cm) {\small$J_t$};
				\node[left=0.25cm] at (0,0) {$P_L$};  	\node[right=0.25cm] at (0,0) {$P_R$}; 	
		\end{tikzpicture}
		\end{minipage}
	}
	\subfloat[]{
		\begin{minipage}[c]{0.24\linewidth}
		\centering
	    \begin{tikzpicture}
				\draw (0,0) circle[radius=1.5cm];
				\draw[orange] (90:1.5cm) -- (270:1.5cm); 
				\draw[orange] (90:1.5cm) -- (140:1.5cm);  \draw[orange] (140:1.5cm) -- (190:1.5cm); 
				\node[right] at (205:1.5cm) {$\textcolor{orange}{\ldots}$};
				\node[above] at (115:1.5cm) {\small$I_1$}; \node[left] at (165:1.4cm) {\small$I_2$};
				\node[below] at (245:1.5cm) {\small$I_s$};
				\draw[orange] (220:1.5cm) -- (270:1.5cm); 
	            \draw (90:1.5cm) -- (40:1.5cm); 
				\draw (40:1.5cm) -- (350:1.5cm);  \draw (320:1.5cm) -- (270:1.5cm); 
				\node[left] at (335:1.5cm) {$\ldots$};
				\node[above] at (65:1.5cm) {\small$J_1$}; \node[right] at (15:1.4cm) {\small$J_2$};
				\node[below] at (295:1.5cm) {\small$J_t$};
				\node[left=0.25cm] at (0,0) {$P_L$};  	\node[right=0.25cm] at (0,0) {$P_R$}; 	
    	\end{tikzpicture}
		\end{minipage}
	}
	\subfloat[]{
		\begin{minipage}[c]{0.24\linewidth}
		\centering
	    \begin{tikzpicture}
			    \draw (0,0) circle[radius=1.5cm];
				\draw[orange] (90:1.5cm) -- (270:1.5cm); 
				\draw (90:1.5cm) -- (140:1.5cm);  \draw (140:1.5cm) -- (190:1.5cm); 
				\node[right] at (205:1.5cm) {$\ldots$};
				\node[above] at (115:1.5cm) {\small$I_1$}; \node[left] at (165:1.4cm) {\small$I_2$};
				\node[below] at (245:1.5cm) {\small$I_s$};
				\draw (220:1.5cm) -- (270:1.5cm); 
	            \draw[orange] (90:1.5cm) -- (40:1.5cm); 
				\draw[orange] (40:1.5cm) -- (350:1.5cm);  \draw[orange] (320:1.5cm) -- (270:1.5cm); 
				\node[left] at (335:1.5cm) {$\textcolor{orange}{\ldots}$};
				\node[above] at (65:1.5cm) {\small$J_1$}; \node[right] at (15:1.4cm) {\small$J_2$};
				\node[below] at (295:1.5cm) {\small$J_t$};
				\node[left=0.25cm] at (0,0) {$P_L$};  	\node[right=0.25cm] at (0,0) {$P_R$}; 	
    	\end{tikzpicture}
		\end{minipage}
	}
  \caption{An illustrative figure for showing the equivalence between different cutting out processes. The position of the ``final block" is indicated by orange.}
  \label{fig:consistency1} 
\end{figure}
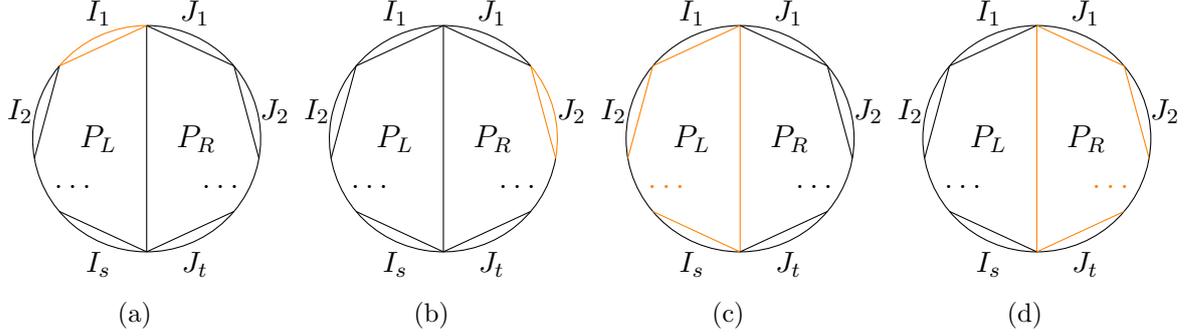
\begin{itemize}
\item \textbf{Case (a)} We can put the final block in any of the $I_{m}$ with $m=1,\ldots,s$, so that the ordering of cutting out $P_{L}$ and $P_{R}$ must be first $P_{R}$ then $P_{L}$.  
\item \textbf{Case (b)} Likewise, we can put the final block in $J_{m}$ with $m=1,\ldots,t$, so that the ordering of cutting out $P_{L}$ and $P_{R}$ must be first $P_{L}$ then $P_{R}$.  
\item \textbf{Case (c)} $P_{L}$ is the final block.
\item \textbf{Case (d)} $P_{R}$ is the final block. 
\end{itemize}
In the former two cases, where neither $P_{L}$ nor $P_{R}$ is the final block, changing the final block, for example from a polygon within  $I_1$ to one within $I_2$  does not change the contribution from cutting out $P_{L}$ and $P_{R}$ so that we put the discussion of these two cases aside. Thus, it is suffice to show that the Cases (c) and (d) are equivalent. To do that, one just explicitly writes down the contributions from $P_{L}$ and $P_{R}$ in Cases (c) and (d). 

\noindent{Case (c)}:
{\small
\begin{equation}
\begin{aligned}
	& \sum_{a=s}^{2} A(I_1,\ldots,I_{a-1},\downarrow,I_a,\ldots,I_{s},J_1,\ldots,J_{t}|I_1^\prime, { \color{blue} \bm{ \downarrow \ldots,I_{a-1}^\prime,\downarrow}},I_a^\prime,\ldots,I_{s}^\prime,(J_1^\prime,\ldots,J_{t}^\prime)^{-1})\\
	&-\sum_{b=2}^{t{-}1} A(I_{s},J_1,\ldots,\downarrow,J_{b},\ldots,J_{t},I_1,\ldots|I_s^\prime, (J_1^\prime,\ldots,{ \color{blue} \bm{ \downarrow, J_b^\prime \ldots,J_{t{-}1}^\prime,\downarrow}},J_{t}^\prime)^{-1},I_1^\prime,\ldots)\\
	&-A(I_{s},\downarrow,J_{1},\ldots,J_{t},I_1,\ldots|I_s^\prime, ({ \color{blue} \bm{ \downarrow, J_1^\prime \ldots,J_{t{-}1}^\prime,\downarrow}},J_{t}^\prime)^{-1},I_1,\ldots)
\end{aligned}
\end{equation}}
and {Case (d)}:
{\small
\begin{equation}
\begin{aligned}
	&  A(I_1,\ldots,I_{s},\downarrow,J_1,\ldots,J_{t}|I_1^\prime, { \color{blue} \bm{ \downarrow \ldots,I_{s}^\prime,\downarrow}},(J_1^\prime,\ldots,J_{t}^\prime)^{-1})\\
	& +\sum_{a=s}^{2} A(I_1,\ldots,I_{a-1},\downarrow,I_{a},\ldots,I_{s},J_1,\ldots,J_{t}|I_1^\prime, { \color{blue} \bm{ \downarrow \ldots,I_{a-1}^\prime,\downarrow}},I_a^\prime,\ldots,I_{s}^\prime,(J_1^\prime,\ldots,J_{t}^\prime)^{-1})\\
	&-\sum_{b=2}^{t{-}1} A(I_{s},J_1,\ldots,J_{b-1},\downarrow,J_{b},\ldots,J_{t},I_1,\ldots|I_s^\prime, (J_1^\prime,\ldots,{ \color{blue} \bm{ \downarrow, J_b^\prime \ldots,J_{t{-}1}^\prime,\downarrow}},J_{t}^\prime)^{-1},I_1^\prime,\ldots)\,.
\end{aligned}
\end{equation}}
Comparing these two equations, it is easy to find out that the difference is 
\begin{equation}\label{eq:phi2temp3}
\begin{aligned}
\text{(d)}-\text{(c)}=&A(I_1,\ldots,I_{s},\downarrow,J_1,\ldots,J_{t}|I_1^\prime, { \color{blue} \bm{ \downarrow \ldots,I_{s}^\prime,\downarrow}},(J_1^\prime,\ldots,J_{t}^\prime)^{-1})\\
&+A(I_{s},\downarrow,J_{1},\ldots,J_{t},I_1,\ldots|I_s^\prime, ({ \color{blue} \bm{ \downarrow, J_1^\prime \ldots,J_{t{-}1}^\prime,\downarrow}},J_{t}^\prime)^{-1},I_1^\prime,\ldots)\,,
\end{aligned}
\end{equation}
and via KK-relation, one gets
\begin{equation}
\begin{aligned}
\text{(d)}-\text{(c)}=&-A(I_1,\ldots,I_{s},\downarrow,J_1,\ldots,J_{t}|I_1^\prime,\ldots,I_s^\prime,(J_1^\prime,\ldots,{ \color{blue} \bm{J_{t}^\prime}})^{-1})\\
&-A(I_1,\ldots,I_{s},\downarrow,J_1,\ldots,J_{t}|{ \color{blue} \bm{I_1^\prime}},\ldots,I_s^\prime,(J_1^\prime,\ldots,J_{t}^\prime)^{-1})\,,
\end{aligned}
\end{equation}
which is zero because it is impossible to draw a mutual partial triangulation for both these two sets of $\phi^3$ amplitudes. Also, we have run a large number of tests that such a statement is correct. In conclusion, the Cases (c) and (d) give equivalent contributions so that one can use the conclusion inductively and prove that any ordering of applying our algorithm leads to the same result. 

\subsection{Unitarity and factorizations}\label{app:unitarity}
Now we move on to prove the expansion for general pure scalar cases by showing that the expansion has the correct factorization behaviour. 

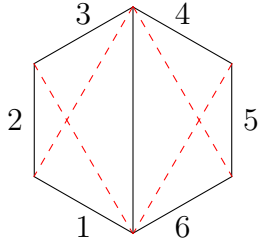
\begin{figure}[htbp!]
  \centering
  \begin{tikzpicture}
    \node[regular polygon,
	draw,minimum size=3cm,
	regular polygon sides = 6,rotate=360/12] (p) at (0,0) {};
	\node[above] at (p.side 1) {3}; \node[left] at (p.side 2) {2};
	\node[below] at (p.side 3) {1};  \node[below] at (p.side 4) {6};
	\node[right] at (p.side 5) {5};  \node[above] at (p.side 6) {4};
	\draw (p.corner 4) -- (p.corner 1); 
	\draw[red,dashed] (p.corner 6) -- (p.corner 4); \draw[red,dashed] (p.corner 2) -- (p.corner 4);
	\draw[red,dashed] (p.corner 1) -- (p.corner 3); \draw[red,dashed] (p.corner 1) -- (p.corner 5);
	\end{tikzpicture}
  \caption{Factorization channels for $F_6^{\mathrm{tr}(\phi^2)}(1,2,3,6,5,4|1,2,3,4,5,6)$, the black lines represent the mutual partial triangulation and each red dashed line represents a diagonal allowed to be added. Every diagonal represents two factorization channels, {\it e.g.} the black one stands for $s_{1,2,3}$ and $s_{4,5,6}$.}
  \label{fig:channel recognization}
\end{figure}

To begin with, we first introduce an efficient way to recognize all of the possible factorization channels for the form factor $F_r^{\operatorname{tr}(\phi^2)}(\alpha|\beta)$ from the mutual partial triangulation related to $(\alpha|\beta)$. The factorization channels are simply represented by one existing diagonal of the polygon or any possible diagonal allowed to be added.  Note that each one of such a diagonal line represents two ``complementary'' factorization channels $s_{I}$ and $s_{\bar{I}}$ with $I\bigcup \bar{I}=(1,2,\ldots,r)$.
For example, the  possible factorization channels of $F_6^{\mathrm{tr}(\phi^2)}(1,2,3,6,5,4|1,2,3,4,5,6)$ given by Figure~\ref{fig:channel recognization} are $s_{1,2,3},s_{4,5,6},s_{1,2},s_{3,4,5,6},s_{2,3},s_{1,4,5,6},s_{4,5},s_{1,2,3,6},s_{5,6},s_{1,2,3,4}$.

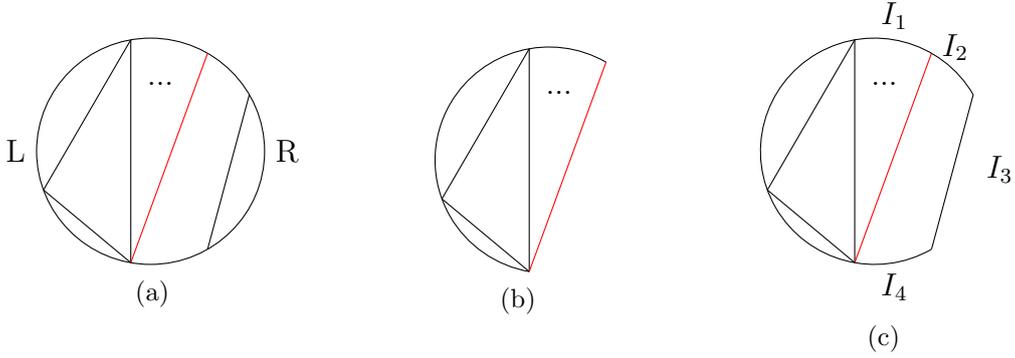
\begin{figure}[htbp!]
	\centering
	\subfloat[]{
		\begin{minipage}[c]{0.3\linewidth}
			\centering
			\begin{tikzpicture}
				\coordinate (p) at (0,0);
				\draw (p) circle[radius=1.5cm];
				\draw (100:1.5cm) -- (200:1.5cm); \draw (260:1.5cm) -- (200:1.5cm);
				\draw (100:1.5cm) -- (260:1.5cm); \draw[red] (60:1.5cm) -- (260:1.5cm);
				\draw (30:1.5cm) -- (300:1.5cm); 
				\node[above] at (80:0.75cm) {...};
				\node[left] at (180:1.5cm) {L};
				\node[right] at (0:1.5cm) {R};
			\end{tikzpicture}
		\end{minipage}
	}
	\subfloat[]{
		\begin{minipage}[c]{0.3\linewidth}
			\centering
			\begin{tikzpicture}
			\coordinate (p) at (0,0);
			\draw (100:1.5cm) -- (200:1.5cm); \draw (260:1.5cm) -- (200:1.5cm);
			\draw (100:1.5cm) -- (260:1.5cm); 
			\draw[red] (0,0) (60:1.5cm) -- (260:1.5cm);
			\node[above] at (80:0.75cm) {...};
			\draw (60:1.5cm) arc(60:260:1.5cm) ; 
		\end{tikzpicture}
		\end{minipage}
	}
	\subfloat[]{
		\begin{minipage}[c]{0.3\linewidth}
		\centering
	\begin{tikzpicture}
		\coordinate (p) at (0,0);
		\draw (100:1.5cm) -- (200:1.5cm); \draw (260:1.5cm) -- (200:1.5cm);
		\draw (100:1.5cm) -- (260:1.5cm); 
		\draw[red] (0,0) (60:1.5cm) -- (260:1.5cm);
		\draw  (30:1.5cm) -- (300:1.5cm); 
		\node[above] at (80:0.75cm) {...};
		\draw (30:1.5cm) arc(30:300:1.5cm) ; 
		\node[above] at (80:1.5cm) {$I_1$}; \node[above] at (45:1.5cm) {$I_2$};
		\node[right] at (350:1.35cm) {$I_3$}; \node[below] at (280:1.5cm) {$I_4$};
	\end{tikzpicture}
		\end{minipage}
	}
  \caption{(a): The mutual partial triangulation for a form factor $F_r^{\operatorname{tr}(\phi^2)}(L,R|L^{\prime},R^{\prime})$. The factorization channel $s_{R}$ is denoted by red line and the edges left(right) to the red line is represented by $L(R)$. (b): A lower point mutual partial triangulation for $F_{|L|+1}^{\mathrm{tr}(\phi^2)}(I^L,L|I^L,L^{\prime})$. (c): An intermediate step for the expansion of original form factor that ``contains'' (b).}
  \label{proof by factorization} 
\end{figure}

With this knowledge, as well as the equivalence proved in \ref{app:equivalencea} and \ref{app:equivalence}, we would like to show that our expansion has the correct factorization behavior: 
consider a mutual partial triangulation in Figure~\ref{proof by factorization}(a) and the factorization channel $s_{R} \rightarrow 0$ represented by the red line therein. 
The edges left(right) to the red line in Figure~\ref{proof by factorization}(a) are denoted by $L(R)$ and their permutation according to $\beta$ ordering are  $L^\prime(R^\prime)$.
The double ordered form factor $F_r^{\operatorname{tr}(\phi^2)}(L,R|L^{\prime},R^{\prime})$  factorizes as
\begin{equation}\label{eq:phi2temp2}
\text{Res}[F_r^{\operatorname{tr}(\phi^2)}(L,R|L^{\prime},R^{\prime})]_{s_{R}=0}= F_{|L|+1}^{\mathrm{tr}(\phi^2)}(I^L,L|I^L,L^{\prime}) A(R,I^R|R^{\prime},I^R) ,
\end{equation}
where the interchanging particle is $I^L,I^R$.

Meanwhile, we choose a construction path cutting out the edges on the right side of the red line first, and get the following sketch expansion  as an intermediate step
\begin{equation}\label{eq:phi2temp4}
    F_r^{\operatorname{tr}(\phi^2)}(L,R|L^{\prime},R^{\prime})= \widetilde{F}^{\operatorname{tr}(\phi^2)}+\sum A_{r{+1}}\,,
\end{equation}
in which the mutual partial triangulation of $\widetilde{F}^{\operatorname{tr}(\phi^2)}$ always ``contains'' Figure~\ref{proof by factorization}(b) as a sub-diagram. Concretely, we can specify the mutual partial triangulation of $\widetilde{F}^{\operatorname{tr}(\phi^2)}$ as Figure~\ref{proof by factorization}(c) where for simplicity we have assumed (c) is obtained from (a) via a single reduction and it is easy to generalize to other cases.
As a result, the RHS of  \eqref{eq:phi2temp4} now becomes
\begin{equation}\label{eq:phi2temp5}
 \widetilde{F}_{n-|I_3|+1}^{\mathrm{tr}(\phi^2)}(\underbrace{\kappa,[I_1}_L,\underbrace{I_2],I_3,[I_4]}_R|\underbrace{\kappa^\prime,[I_1^\prime}_{L^\prime},\underbrace{I_2^\prime],I_3^\prime,[I_4^\prime]}_{R^\prime}) + \sum A_{r+1}(\rho,{\color{blue}\bm I_3}|\rho^{\prime},{\color{blue}\bm I_3^{\prime}}).
\end{equation}
where $\kappa$ represents the particles aside from those in $I_1,I_2,I_3,I_4$ in the corresponding ordering and similar for $\kappa^\prime$ and $\rho,\rho^{\prime}$. We use the square brackets to emphasize the particles inside should not be regarded as a subgroup $I_i$, say $\widetilde{F}^{\operatorname{tr}(\phi^2)}_{n{-}|I_3|{+}1}$ is not necessarily proportional to $s_{I_i}$ for $i=1,2,4$. 
Note that the $\sum A_{r{+}1}$ does not contribute to the factorization channel $s_{R} \rightarrow 0$, since they are bi-adjoint amplitudes with $q$ inserted inside $I_3$. 

Next, we use the following fact
\begin{equation}
\begin{aligned}
    &\widetilde{F}_{n-|I_3|+1}^{\mathrm{tr}(\phi^2)}(\kappa,[I_1,I_2],I_3,[I_4]| \kappa^\prime,[I_1^\prime,I_2^\prime],I_3^\prime,[I_4^\prime])\\
    =& A([I_3],I^+|[I_3^\prime],I^+) \frac{1}{s_{I_3}} F_{n-|I_3|+1}^{\mathrm{tr}(\phi^2)}(\kappa,[I_1,I_2],I^-,[I_4]|\kappa^\prime,[I_1^\prime,I_2^\prime],I^-,[I_4^\prime]),
\end{aligned}
\end{equation}
and take the residue of \eqref{eq:phi2temp5} on the $s_{R}=0$ pole, getting 
\begin{equation}
\text{Res}[\eqref{eq:phi2temp5}]_{s_{R}=0}=
    A([I_3],I^+|[I_3^\prime],I^+) \frac{1}{s_{I_3}} F_{|L|+1}^{\mathrm{tr}(\phi^2)}(I^L,L|I^L,L^{\prime}) A(\widetilde{R},I^R|\widetilde{R^{\prime}},I^R),
\end{equation}
where $\widetilde{R}$ represent edges right to the red line but $I_3$ is regarded as a single particle and similar for $\widetilde{R^{\prime}}$. Importantly, the $\phi^3$ amplitudes appear above satisfy
\begin{equation}
    A([I_3],I^+|[I_3^\prime],I^+) \frac{1}{s_{I_3}} A(\widetilde{R},I^R|\widetilde{R^{\prime}},I^R)=A(R,I^R|R^{\prime},I^R).
\end{equation}
At the end of the day, we obtain 
\begin{equation}
  \text{Res}[\eqref{eq:phi2temp5}]_{s_{R}=0}=F_{|L|+1}^{\mathrm{tr}(\phi^2)}(I^L,L|I^L,L^{\prime}) A(R,I^R|R^{\prime},I^R) ,
\end{equation}
which is exactly $\text{Res}[F_r^{\operatorname{tr}(\phi^2)}(L,R|L^{\prime},R^{\prime})]_{s_{R}=0}$ in \eqref{eq:phi2temp2}. This completes our proof. 

\subsection{Factorizations for ${\rm tr}(\phi^2)$ form factor with gluons} \label{app:proof general gluon}
\begin{figure}[htbp!]
	\centering
	\begin{tikzpicture}
		\node[regular polygon,
		draw,minimum size=3cm,
		regular polygon sides = 6,rotate=360/12] (p) at (0,0) {};
		\node[above] at (p.side 1) {3}; \node[left] at (p.side 2) {2};
		\node[below] at (p.side 3) {1};  \node[below] at (p.side 4) {6};
		\node[right] at (p.side 5) {5};  \node[above] at (p.side 6) {4};
		\draw (p.corner 4) -- (p.corner 1); 
		\draw[red,dashed] (p.corner 6) -- (p.corner 4); \draw[red,dashed] (p.corner 2) -- (p.corner 4);
		\draw[red,dashed] (p.corner 1) -- (p.corner 3); \draw[red,dashed] (p.corner 1) -- (p.corner 5);
		\fill[gray!50] (p.corner 4) circle (3pt);
		\node[below] at (p.corner 4) {$\mathcal{G}_{6}=(7,8)$};
	\end{tikzpicture}
	\caption{Factorization channels for $F_8^{\mathrm{tr}(\phi^2)}(1,2,3,6,5,4|1,2,3,4,5,6,7,8)$. The black diagonal represent $s_{4,5,6},s_{4,5,6,7},s_{4,5,6,7,8}$ and their complementary channels.}
	\label{fig:channel recognization gluon}
\end{figure}
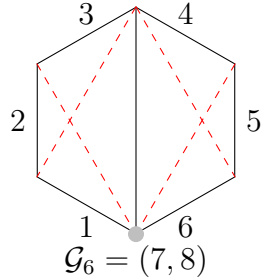

As we have illustrated, the expansion for form factors with gluons is trivially given once its scalar skeleton expansion is done. In fact, the consistency check of factorizations in this case is a simple generalization of the pure scalar case. Now we  generalize the procedure for producing all possible factorization channels with scalar propagator(gluon propagator cases are trivial): 

For the form factor $F_n^{{\rm tr}(\phi^2)}(\alpha(i_1,i_2,\ldots,i_r)| i_1,\mathcal{G}_{i_1},i_2,\mathcal{G}_{i_2},\ldots,i_r,\mathcal{G}_{i_r})$, we draw the mutual partial triangulation for its scalar skeleton $(i_1,i_2,\ldots,i_r|\alpha(i_1,i_2,\ldots,i_r))$ with edges labeled by $i_1,i_2,\ldots,i_r$. Importantly, the vertex between edges $i_a,i_{a{+}1}$ is now labeled by $\mathcal{G}_{i_a}$, {\it i.e.} the gluons inserting between these scalars. Then, similar to the pure scalar case, the diagonal of the polygon and any possible diagonal allowed to be added represents possible factorization channels. 
The difference is that now such a diagonal stands not only for two channels but more, {\it e.g.} the diagonal that connecting vertex $\mathcal{G}_{i_c}$ and $\mathcal{G}_{i_d}$ is now representing channels(or their complementary channels) $s_{\mathcal{G}_{i_{c}}^\prime,i_{c+1}, \mathcal{G}_{i_{c+1}},\ldots,i_d, \mathcal{G}_{i_{d}}^\prime} \rightarrow 0$ with $\mathcal{G}_{i_{c}}^\prime, \mathcal{G}_{i_{d}}^\prime$ to be ordered subsets of $\mathcal{G}_{i_{c}},\mathcal{G}_{i_{d}}$ adjacent to $i_{c+1}, i_d$ respectively. An explicit example for $F_8^{{\rm tr}(\phi^2)}(1,2,3,6,5,4|1,2,3,4,5,6,7,8)$ is given in Figure \ref{fig:channel recognization gluon}. 

With the above description of the possible factorization channel, the unitarity analysis can be easily generalized from the pure scalar discussions.

\section{Review of CHY formula}  \label{app:CHY}
Cachazo-He-Yuan fomulae~\cite{CHY1,CHY2,CHY3} express tree-level scattering amplitudes of a wide range of theories in arbitrary dimension as an integral over the moduli space of genus zero Riemann surfaces with $n$ punctures. The concise expression can be summarized by:
\begin{equation}
	A_{n}=\int d \mu_{n} \mathcal{I}_{L}(\{p, \epsilon, \sigma\}) \mathcal{I}_{R}(\{p, \tilde{\epsilon}, \sigma\}),
\end{equation}
where $\sigma_a$ is the puncture location for $a$th particle and the mesure is defined to be:
\begin{equation}
	\begin{aligned}
		d \mu_{n} & := \sigma_{i j} \sigma_{j k} \sigma_{k i} \prod_{a \neq i, j, k} \delta(E_{a}) \frac{d^{n} \sigma}{\operatorname{vol\ SL}(2, \mathbb{C})} \\
		&=\left(\sigma_{i j} \sigma_{j k} \sigma_{k i}\right)\left(\sigma_{r t} \sigma_{t s} \sigma_{s r}\right) \prod_{a \neq i, j, k} \delta(E_{a}) \prod_{b \neq r, t, s} d \sigma_{b},
	\end{aligned}
\end{equation}
here we fix three punctures $\sigma_r, \sigma_t,\sigma_s$ by the ${\rm SL}(2,\mathbb{C})$ invariance. We also define $\sigma_{ab}:= \sigma_a-\sigma_b$
and the scattering equations are given by
\begin{equation}
	E_{a}=\sum_{b \neq a} \frac{p_a \cdot p_b}{\sigma_{a b}}.
\end{equation}
For our purpose, we now introduce CHY integrand for YMS amplitudes~\cite{CHY4}. The first building block is the Parke-Taylor factor:
\begin{equation}
	\mathrm{PT}(\alpha)=\frac{1}{\sigma_{\alpha_1\alpha_2} \sigma_{\alpha_2 \alpha_3} \ldots \sigma_{\alpha_{r{-}1} \alpha_r}\sigma_{\alpha_r \alpha_1}},
\end{equation}
where $|\alpha|=r$. For another ingredient, we first introduce the $2n \times 2n$ matrix $\Psi$:
\begin{equation}
	\Psi=\left(\begin{array}{cc}
		A & -C^{T} \\
		C & B
	\end{array}\right),
\end{equation}
with the $n \times n$ matrices $A,B,C$ defined by
\begin{equation}
	A_{a b}=\left\{\begin{array}{ccl}
		\frac{p_{a} \cdot p_{b}}{\sigma_{a b}} & a \neq b \\
		0 & a=b
	\end{array} \quad B_{a b}=\left\{\begin{array}{cc}
		\frac{\epsilon_{a} \cdot \epsilon_{b}}{\sigma_{a b}} & a \neq b \\
		0 & a=b
	\end{array} \quad C_{a b}=\left\{\begin{array}{cc}
		\frac{\epsilon_{a} \cdot p_{b}}{\sigma_{a b}} & a \neq b \\
		-\sum_{c \neq a} \frac{\epsilon_{a} \cdot p_{c}}{\sigma_{a c}} & a=b
	\end{array}\right.\right.\right.
\end{equation}
for $1\leq a,b,c \leq n$. Precisely, we need the submatrix 
\begin{equation}
	\Psi({\mathrm{G}})=\left(\begin{array}{cc}
		A_{\mathrm{G}} & -\left(C_{\mathrm{G}}\right)^{T} \\
		C_{\mathrm{G}} & B_{\mathrm{G}}
	\end{array}\right),
\end{equation}
where $\mathrm{G}=\{g_1,g_2,\ldots,g_m\}$  represents the set of gluons. The matrices $A_{\mathrm{G}},B_{\mathrm{G}},C_{\mathrm{G}}$ are submatrices of $A,B,C$ whose row and column indices take value only in $\mathrm{G}$. The YMS amplitudes under the CHY formula is then given by:
\begin{equation}
	A(\alpha|1,2,\ldots,n)= (-1)^{\frac{m(m{+}1)}{2}+m} \int d \mu_{n} \mathrm{PT}(1,2,\ldots,n) \mathrm{PT}(\alpha) \operatorname{Pf} \Psi(\mathrm{G}),
\end{equation}
where $\alpha$ is the complementary set of $\mathrm{G}$ under a given order, for $m=0$ the expression computes amplitudes for bi-adjoint $\phi^3$ theory. Here we use the sign convention according to~\cite{Du:2017kpo}. For notation convenience, we absorb the sign by defining
\begin{equation}
    \operatorname{Pf} \Psi_m := (-1)^{\frac{m(m{+}1)}{2}+m} \operatorname{Pf} \Psi(\mathrm{G}).
\end{equation}

\providecommand{\href}[2]{#2}\begingroup\raggedright\endgroup

\end{document}